\documentstyle[11pt,aaspp4,flushrt]{article}

\def\spose#1{\hbox to 0pt{#1\hss}}

\def\kms{\ifmmode {\rm\,km\,s^{-1}}\else
    ${\rm\,km\,s^{-1}}$\fi}
\def\kmsMpc{\ifmmode {\rm\,km\,s^{-1}\,Mpc^{-1}}\else
    ${\rm\,km\,s^{-1}\,Mpc^{-1}}$\fi}

\def\msun{\ifmmode {\rm\,M_\odot}\else ${\rm\,M_\odot}$\fi}
\def\Msun{\ifmmode {\rm\,M_\odot}\else ${\rm\,M_\odot}$\fi}
\def\lsun{\ifmmode {\rm\,L_\odot}\else ${\rm\,L_\odot}$\fi}
\def\Lsun{\ifmmode {\rm\,L_\odot}\else ${\rm\,L_\odot}$\fi}
\def\rsun{\ifmmode {\rm\,R_\odot}\else ${\rm\,R_\odot}$\fi}
\def\Rsun{\ifmmode {\rm\,R_\odot}\else ${\rm\,R_\odot}$\fi}

\def\cm{{\rm\,cm}}
\def\cm3{\ifmmode {\rm\,cm^{-3}}\else ${\rm\,cm^{-3}}$\fi}

\def\ergps{\ifmmode {\rm\,erg\,s^{-1}}\else ${\rm\,erg\,s^{-1}}$\fi}
\def\ergpscm2{\ifmmode {\rm\,erg\,s^{-1}\,cm^{-2}}\else
    ${\rm\,erg\,s^{-1}\,cm^{-2}}$\fi}

\def\cf{{cf.~}\ }
\def\eg{{e.g.}}
\def\deg{\ifmmode {^{\circ}}\else {$^\circ$}\fi}
\def\degr{\ifmmode {^{\circ}}\else {$^\circ$}\fi}
\def\degs{\ifmmode {^{\circ}}\else {$^\circ$}\fi}

\def\etal{{et al.~}}

\def\h3Mpc{h^{-3}{\rm Mpc}^3}
\def\Ho{\ifmmode {H_0}\else ${H_0}$\fi}
\def\hnot{\ifmmode {H_0}\else ${H_0}$\fi}
\def\h0{\ifmmode {H_0}\else ${H_0}$\fi}
\def\hnotunit{\ifmmode {\rm\,km\,s^{-1}\,Mpc^{-1}}\else
    ${\rm\,km\,s^{-1}\,Mpc^{-1}}$\fi}
\def\qnot{\ifmmode {q_0}\else ${q_0}$\fi}
\def\q0{\ifmmode {q_0}\else ${q_0}$\fi}
\def\ie{{i.e.}}
\def\mic{\ifmmode {\rm\,\mu m}\else ${\rm \mu m}$\fi}

\def\arcsec{\ifmmode {^{\prime\prime}}\else $^{\prime\prime}$\fi}
\def\asec{\ifmmode {^{\prime\prime}}\else $^{\prime\prime}$\fi}
\def\arcmin{\ifmmode {^{\prime}}\else $^{\prime}$\fi}
\def\amin{\ifmmode {^{\prime}}\else $^{\prime}$\fi}

\def\secper{\ifmmode \rlap.{^{s}}\else $\rlap{.}{^{s}} $\fi}
\def\minper{\ifmmode \rlap.{^{m}}\else $\rlap{.}{^m} $\fi}
\def\magper{\ifmmode \rlap.{^{m}}\else $\rlap{.}{^m} $\fi}
\def\farcs{\ifmmode \rlap.{^{\prime\prime}}\else
    $\rlap.{^{\prime\prime}}$\fi}
\def\arcsper{\ifmmode \rlap.{^{\prime\prime}}\else
    $\rlap.{^{\prime\prime}}$\fi}
\def\arcmper{\ifmmode \rlap.{^{\prime}}\else
    $\rlap.{^{\prime}}$\fi}
\def\spose#1{\hbox to 0pt{#1\hss}}
\def\simlt{\mathrel{\spose{\lower 3pt\hbox{$\mathchar"218$}}
     \raise 2.0pt\hbox{$\mathchar"13C$}}}
\def\simgt{\mathrel{\spose{\lower 3pt\hbox{$\mathchar"218$}}
     \raise 2.0pt\hbox{$\mathchar"13E$}}}

\def\araa{{ARA\&A}}
\def\aa{{A\&A}}

\def\aasupp{{A\&AS}}
\def\aj{{AJ}}
\def\apj{{ApJ}}

\def\apjsupp{{ApJS}}
\def\apjs{{ApJS}}

\def\mn{{MNRAS}}
\def\mnras{{MNRAS}}
\def\nature{{Nature}}

\def\pasp{{PASP}}

\def\apjref#1;#2;#3;#4 {\par\pp#1, {#2}, #3, #4 \par}

\tighten

\slugcomment{\it Accepted for publication in The Astrophysical Journal} 
 
\lefthead{Spinrad et al.}
\righthead{An Old Galaxy at High Redshift}

\begin{document}

\title{LBDS~53W091: An Old, Red Galaxy at z=1.552\altaffilmark{1}}

\author{Hyron Spinrad} 
\affil{Astronomy Department, University of California at Berkeley, CA 94720}
\affil{Electronic Mail: spinrad@astro.berkeley.edu}

\author{Arjun Dey}
\affil{NOAO/KPNO\altaffilmark{2}, 950 N. Cherry Ave., P. O. Box 26732, Tucson, AZ 85726}
\affil{Electronic Mail: dey@noao.edu}

\author{Daniel Stern}
\affil{Astronomy Department, University of California at Berkeley, CA 94720}
\affil{Electronic Mail: dan@astro.berkeley.edu}

\author{James Dunlop}
\affil{Institute for Astronomy, Department of Physics and Astronomy} 
\affil{The University of Edinburgh, Edinburgh EH9 3HJ, UK}
\affil{Electronic Mail: J.Dunlop@roe.ac.uk}

\author{John Peacock and Raul Jimenez}
\affil{Royal Observatory, Edinburgh EH9 3HJ,  UK}
\affil{Electronic Mail: (J.Peacock,R.Jimenez)@roe.ac.uk}

\author{Rogier Windhorst}
\affil{Department of Physics and Astronomy, Arizona State University, Tempe,
AZ 85287-1504}
\affil{Electronic Mail: raw@cosmos.la.asu.edu}

\altaffiltext{1}{Based in large part 
on observations made at the W.M. Keck Observatory.}
\altaffiltext{2}{The National Optical Astronomy Observatories are
operated by the Association of Universities for Research in Astronomy under
cooperative agreement with the National Science Foundation.}

\begin{abstract}

The weak radio source LBDS~53W091 is associated with a very faint ($R
\approx 24.5$) red ($R-K \approx 5.8$) galaxy. Long spectroscopic
integrations with the W.\ M.\ Keck telescope have provided an
absorption--line redshift, $z=1.552 \pm 0.002.$ The galaxy has a rest
frame ultraviolet spectrum very similar to that of an F6~V star, and a
single--burst old stellar population that matches the IR colors, the
optical energy distribution and the spectral discontinuities has a
minimum age of 3.5 Gyr. We present detailed population synthesis
analyses of the observed spectrum in order to estimate the time since
the last major epoch of star formation. We discuss the discrepancies in
these estimates resulting from using different models, subjecting the
UV spectrum of M32 to the same tests as a measure of robustness of
these techniques.  The models most consistent with the data tend to
yield ages at $z=1.55$ of $\simgt 3.5$~Gyr, similar to that inferred
for the intermediate--age population in M32.  Depending upon the
assumed Hubble constant and the value of $\Omega_0,$ only certain
cosmological expansion times are consistent with the age of
LBDS~53W091; in particular, for $\Omega_0 = 1,$ only models with $H_0
\lesssim 45$ km s$^{-1}$ Mpc$^{-1}$ are permitted. For $H_0 = 50$ km
s$^{-1}$ Mpc$^{-1}$ and $\Omega_0 = 0.2,$ we derive a formation
redshift, $z_f \ge 5.$

\end{abstract}


\section{Introduction}

Finding distant galaxies and analyzing their starlight remains one of
the only direct methods of studying the formation and evolution of
galaxies. In particular, the reddest normal galaxies at high redshifts provide
the best constraints on the earliest epochs of galaxy formation and
evolution, since their color is most likely due to an aged stellar
population. Several photometric and spectroscopic studies of galaxy
evolution out to redshifts $z\sim1$ have discovered that the red galaxy
population (which predominantly consists of early type E/S0 galaxies)
evolves ``passively'' with time, \ie, by the gradual reddening and
fading of the integrated starlight (\eg, Driver \etal 1995ab;
Lilly \etal 1995; Rakos \& Schombert 1995; Schade \etal 1995;
Stanford \etal 1995, 1997a; Oke \etal 1996). 
In addition, the discovery of $z\sim 1$ cluster galaxies with 
morphologies and rest frame colors similar to those of nearby
ellipticals (\eg, Couch \etal 1994; Dressler \etal 1995; 
Dickinson 1996; Dickinson \etal 1997) 
suggests a high formation redshift ($z>2$) for the red
population and emphasizes the importance of studying these objects at
even larger lookback times.

The high-redshift red galaxy population is faint at observed optical
(rest--frame ultraviolet) wavelengths, and therefore most studies of galaxies at
high redshift have concentrated on the luminous, blue, emission line
objects (star--forming and active galaxies) which are easier to find
and relatively easy to study spectroscopically at optical wavelengths
(\eg, Cowie \etal 1995, Steidel \etal 1996). One of the prerequisites
to studying old populations at high redshifts is therefore to find
distant luminous early type galaxies.  The association of nearby,
bright radio sources with low redshift giant elliptical and cD galaxies
suggests that a good method of finding such old populations at high
redshifts is to search for the optical counterparts of faint radio
sources (Kron \etal 1985).
This has been the primary driving force behind several radio
source identification and redshift determination programs over the last
three decades, and has resulted in several nearly completely
identified radio source catalogues (\eg, 3CR --- Spinrad \& Djorgovski 1987;
Molonglo --- McCarthy \etal 1996; 1Jy --- Lilly 1989;
Parkes --- Dunlop \etal 1989a;
2Jy --- Tadhunter \etal 1993; MG --- Stern \etal 1997). 

Unfortunately, most of these studies, although resulting in a large
number of high-redshift objects, are of limited use for studying the
evolution of normal galaxies. This is primarily because the ultraviolet
(UV) light in most luminous radio galaxies is dominated by scattered light
from and photoionization by the active nucleus rather than starlight
(\eg, McCarthy \etal 1987; Chambers \etal 1987; di Serego Alighieri
\etal 1989; di Serego Alighieri \etal 1994; Jannuzi \& Elston 1991;
Jannuzi \etal 1995; Dey \& Spinrad 1996; Dey \etal 1996; Cimatti \etal
1996).  Nevertheless, there have been several attempts to age-date the
underlying stellar population using broad band optical and
near--infrared photometry (\eg, Dunlop \etal 1989b; 
Chambers \& Charlot 1990; McCarthy 1993)
and a few valiant efforts using moderate signal--to--noise
ratio spectroscopy (\eg, Stockton, Kellogg \& Ridgway 1995; Chambers \&
McCarthy 1990). These attempts have been limited by the inherent
ambiguities of modelling broad band colors and, in the spectroscopic
studies, the problems of subtracting the strong emission lines and UV
non-stellar continuum light and correctly decomposing the AGN and
stellar components.

Although radio galaxies have been, thus far, of limited cosmological
utility, they are not to be discarded as useful probes of the early
epochs of galaxy formation and evolution. First, they are still the
highest redshift galaxy-like objects (\ie, spatially extended and
possibly composed of stars) known (\eg, Lacy \etal 1995; Spinrad, Dey
\& Graham 1995; Rawlings \etal 1996).  Second, there appears to be a
good correlation between radio power and the fractional contribution of
non-stellar AGN light to the UV spectrum; in particular, weak radio
sources ($S_{\rm 1.4~GHz} < 50$ mJy) generally have very weak emission
lines and, unlike the powerful radio galaxies, do not exhibit UV /
radio alignments, suggesting that the contribution of scattered AGN
emission to their continuum light is small (\eg, Rawlings \& Saunders 1991; 
Dunlop \& Peacock 1993; Eales \& Rawlings 1993; Vigotti \etal
1996).  Hence, {\it weak} radio sources with red optical / IR colors
may {\it still} provide us with the ability of studying uncontaminated
starlight in nearly normal, luminous elliptical galaxies at high
redshift.  The radio source selection above a few mJy almost guarantees
an early type host galaxy (\eg, Dunlop, Peacock \& Windhorst 1995 and
references therein) and the near--IR magnitude and color criteria
ensure that the galaxy will be at high redshift. For reference, a
present-day $L^*$ elliptical galaxy observed at a redshift $z\approx
1$, has a typical magnitude of $K\approx18.5$ and color $(R-K)\approx6$.

In order to further test this hypothesis, we have chosen as targets
for deep optical spectroscopy a subset of
weak radio sources (1~mJy $< S_{\rm 1.4~GHz} < 50$~mJy) from the
Leiden-Berkeley Deep Survey (hereinafter LBDS; Windhorst \etal 1984ab)
which are associated with host galaxies that have faint near--IR
magnitudes ($K\ge18$) and red optical-IR colors ($R-K>5$).
Photometry is now available for a statistically complete sample of 77
galaxies having $griJHK$ photometry to $r \simeq 26$ and $K \simeq 20$
(Dunlop, Peacock, \& Windhorst 1995). In this paper, we present our
results on LBDS~53W091, a weak radio source ($S_{\rm 1.4~GHz} \approx
20$~mJy) which is among the reddest faint LBDS galaxies, suggesting a
substantial distance and an aged population.  Early results on
this galaxy have already been reported by us elsewhere (Dunlop \etal
1996), and the present work includes a more detailed description of our
data, spectral analyses, and age-dating techniques. In \S~2 we present
our optical, IR and radio imaging photometry and optical spectroscopy.
The redshift determination is described in \S~3.  The derived age
estimates based on spectral synthesis model fitting and the differences
between the various models are presented in \S 4. In \S~5 we discuss
the cosmological implications of finding such an old galaxy at high
redshift.

\section{Observations}

\subsection{Optical Identification, Radio Imaging and Astrometry}

The optical counterpart of the radio source LBDS~53W091 was first
identified on images of the field obtained using the Palomar 200" Hale
Telescope.  The Four-shooter CCD-array on the Hale Telescope was used
in 1984 -- 1988 to systematically image those
mJy radio sources in the 17$^h$+50$^o$ LBDS field (Windhorst, van
Heerde, \& Katgert 1984; WHK) that were fainter than $V\le 23.5$ mag
(\ie, sources not detected on the deep $UJFN$ plates obtained with the
KPNO 4-m Mayall Telescope; Windhorst, Kron, \& Koo 1984; WKK). The
Four-shooter imaging was done in Gunn $g$ and $r$.  Each frame consists
of four simultaneously exposed 800$\times$800 TI CCDs, and covers
$\approx9^\prime \times 9^\prime$.

Details of the Four-shooter imaging, calibration, and reduction are
given by Neuschaefer \& Windhorst (1995a, b; NW95a, NW95b). This
includes a careful removal of large scale gradients to within 0.1\% of
sky, so that aperture magnitudes could be reliably grown to total (see
Windhorst et al 1991).  Photometric calibration was done measuring
standard stars from Thuan \& Gunn (1976) and Kent (1985), and correcting
for atmospheric extinction as a function of airmass and ($g-r$)
color. From overlapping Four-shooter regions and multiple exposures
during different observing runs, we could check the internal
consistency of the photometry during these runs, which was usually
$\le 0.08 - 0.1$ mag (NW95a). Astrometry was done with typically 30
primary standard stars from recent Palomar 48 inch Schmidt plates, and
6--8 standard stars in each Four-shooter CCD, as described by WKK and
NW95a. With repeated astrometric measurements under different plate
orientations, a global astrometric accuracy could be obtained of
$0\farcs3-0\farcs5$. The Westerbork radio positions of WHK and the VLA
positions of Oort \etal (1987) (with typical accuracies of
$0\farcs2-0\farcs3$) were sufficient to find a reliable optical
identification for each source.

High resolution radio images of LBDS~53W091 at frequencies of 1.56~GHz
and 4.86~GHz were obtained using the VLA A-array in snapshot mode on
1995 October 29. Figure~\ref{vla} shows the 4.86~GHz map of the radio
source, and the radio data are presented in Table~\ref{radiodata}.  The
source is a double--lobed FRII steep--spectrum ($\alpha_{\rm
1.56~GHz}^{\rm 4.86~GHz} \approx 1.1,\ S_\nu \propto \nu^{-\alpha}$)
radio source. The radio lobes are separated by $\approx 4{\arcsper}3$
in position angle ${\rm PA \approx 131^\circ}$.

The VLA A-array position of LBDS~53W091 is RA=17$^h$ 21$^m$
$17{\secper}81\pm0{\secper}01$, DEC=+50$^\circ$ 08$^\prime$
$47{\farcs}4\pm0{\farcs}1$ (B1950; Oort et al. 1987), and the best
astrometric position for the optical candidate for LBDS~53W091 is
RA=17$^h$ 21$^m$ 17{\secper}84$\pm$0{\secper}03, DEC=+50$^\circ$
08$^\prime$ $47\farcs7\pm0\farcs3$ (B1950; NW95a). Its optical fluxes
result in magnitudes in Gunn $g\ge26.0$ (2$\sigma$) and
$r=25.10\pm0.15$ mag (see NW95a,b for details). Its 1.41 GHz radio flux
density is 22.4$\pm$0.9 mJy from WSRT observations (with a beamsize of
12\arcsec\ FWHM) in 1980--1984 (Windhorst et al.  1984, Oort \& van
Langevelde 1987). Its 0.61~GHz WSRT flux density is 66.0$\pm$3.9 mJy,
implying a 0.61--1.41~GHz spectral index of 1.30$\pm$0.13. The source is
resolved at the 1.4\arcsec\ FWHM VLA A-array resolution, and has
LAS=$4\farcs2\pm0\farcs5$ (Oort et al.\ 1987). The 1.490~GHz VLA
A--array flux density measured in 1985, transformed back to 1.41~GHz
with the measured spectral index, was $S_{1.41}$=28.8$\pm$1.5~mJy.  The
1995 VLA A--array flux density, transformed to 1.41 GHz with the
spectral index calculated from those observations, was
$S_{1.41}$=25.9$\pm$1.9~mJy.

The VLA A--array observations were done at $\sim 10\times$ higher
resolution than the WSRT observations, and therefore may systematically
miss flux. It is therefore curious that the 1985 VLA 1.41 GHz flux density is
slightly higher (at the combined 3.7$\sigma$ level) than the 1980-1984
WSRT 1.41 GHz flux density, so that the possibility of {\it weak} nuclear
variability cannot be ruled out. However, given its weak radio flux,
steep--spectrum, and small but resolved angular size, the radio
properties point at best to a relatively weak AGN. We note that the
occurrence of a faint {\it red} identification for a compact
weak radio source is quite common in the LBDS sample (\cf
Kron \etal 1985, Windhorst \etal 1985), but less common in a $\mu$Jy
sample (Windhorst et al. 1995).

\subsection{Optical and Near-Infrared Imaging and Photometry}

We obtained an $R$-band image of the field of LBDS~53W091 using the
Low-Resolution Imaging Spectrometer (LRIS; Oke \etal 1995) on the
W.~M.~Keck Telescope on UT 1995 July 25. The LRIS detector is a
Tektronix $2048^2$ CCD with 24~$\mu$m pixels corresponding to a scale
of 0{\arcsper}214 pixel$^{-1}$.  We obtained two 300$s$ exposures under
photometric conditions in fairly good seeing (the coadded image has
FWHM$_{PSF}\approx 1\arcsec$). The images were bias-corrected and
flat-fielded using a median image of the twilight sky.  Photometric
calibration was performed using observations of the standard field
SA~113 (Landolt 1992).  The coadded Keck $R$ image is shown in
Figure~\ref{chart}, and reaches a 3$\sigma$ limiting magnitude of 25.6
in a 4\arcsec\ diameter aperture. A detail of this image centered on
LBDS~53W091 is shown in Figure~\ref{images}a.

Near-infrared images of LBDS~53W091 were obtained using the 3.9-m
United Kingdom Infrared Telescope (UKIRT). On UT 1993 May 16 we
obtained a 54-minute $K$-band image using the 62 $\times$ 58 pixel InSb
array camera IRCAM1, with the camera operating in the 0.62 arcsec
pixel$^{-1}$ mode.  Deep $J$-band (54 minutes) and $H$-band (81 minutes) images
of LBDS~53W091 were subsequently obtained on UT 1995 August 19 using
the 256 $\times$ 256 pixel InSb array camera IRCAM3, with an image
scale of 0.286 arcsec pixel$^{-1}$.  The infrared images were
constructed from a mosaic of short-exposure ($<$ 3 minutes) frames which
were shifted with respect to each other by between 8 and 15 arcsec.
This procedure meant that the target source fell on a different set of
pixels in each frame, and so the frames could be median filtered to
provide an accurate sky flat-field for the image concerned.  The
reduction procedure was as follows: (i) subtraction of a dark/bias
frame from each sub-image; (ii) removal of known bad pixels; (iii)
scaling of each image to the same median level, followed by median
filtering of the stack; (iv) normalization of the resulting flat field;
(v) division of each sub-image by the flat-field; (vi) construction of
the final mosaic involving accurate registration, subtraction of
frame-to-frame DC variations, and averaging of regions of overlap.  The
resulting mosaiced images reach 3$\sigma$ detection limits of $\mu_K \simeq 21$
mag arcsec$^{-2}$, $\mu_H \simeq 22$ mag arcsec$^{-2}$ and $\mu_J
\simeq 23.5$ mag arcsec$^{-2}$.  A detail of the $J+H$ image is
presented in Figure~\ref{images}b, and the optical and
near--infrared photometry are presented in Table \ref{photometry}.

Figure~\ref{colorpic} (Plate 1) shows a false-color composite of the
field constructed using the $R$, $J$ and $H$ images.  There are three
red compact objects that appear to be in a close group near the center
of the field. LBDS~53W091 is associated with the western--most and
brightest red object in the central triad, and is clearly one of the
reddest objects in the field, with $(R-K)\approx 5.8$ (Table
\ref{photometry}). The two galaxies that lie immediately to the NE and
SE of LBDS~53W091 appear to have similar colors and may be companion
galaxies.  The two blue galaxies that lie near LBDS~53W091 (labelled
``1'' and ``3b'' in Figure~\ref{images}) are both foreground emission
line systems as described below.

Our images show that the three red galaxies are marginally resolved
(seeing deconvolved FWHM $\approx$ 0{\arcsper}5 - 0{\arcsper}7), and
the images are consistent with the galaxies being symmetric. More 
detailed comments on the rest frame UV and optical morphologies 
await observations with the {\it Hubble Space Telescope} ({\it HST}).

\subsection{Spectroscopy}

%
%
%

We observed LBDS~53W091 at the Cassegrain focus of the 10-m W.~M.~Keck
Telescope using LRIS in May, July, August and September 1995.  We used
a 300~line/mm grating ($\lambda_{\rm blaze} = 5000$\AA) to cover the
wavelength region $\lambda\lambda 4000 - 9500$\AA\ and a 1\arcsec\ slit
which resulted in a resolution ${\rm FWHM \approx 10\AA.}$ The data
from UT 1995 July 25, August 31 and September 1 were of the best
quality: the galaxy was detected in all these individual spectra and
the seeing varied between 0\arcsper8 and 1\arcsper0 during the
observations. These observations were all made with the slit oriented
at position angle $PA=126^\circ$ in order to obtain spectra of the two
brightest red objects in the field, LBDS~53W091 and galaxy 3a (\eg,
Dunlop \etal 1996).  On these nights, the parallactic angle varied
between $95^\circ$ and $150^\circ$, and our relative spectrophotometry
should not be adversely affected by atmospheric refraction.

The data were bias-corrected, and flat-fielded using internal quartz
flats obtained immediately after each observation. These observations
of LBDS~53W091, galaxy 3$a$ and 3$b$ (see Figure~\ref{images} for
nomenclature) were extracted using apertures of 1{\arcsper}7 (8
pixels).  The individual spectra were wavelength calibrated using HgKr
and NeA lamps obtained after each observation.  Flux calibration was
performed using, on different nights, observations of the standard
stars Feige 110, BD+33$^\circ$2642, G191B2B and Wolf~1346. Standard
star spectra were obtained both with and without a GG495 order-blocking
filter in order to correct for the second order light at long
wavelengths. The flux calibrated spectra of LBDS~53W091 from different
nights are consistent with each other: the average flux in the
wavelength region from 6500\AA\ to 8500\AA\ showed night-to-night
variations of less than 15\%. Finally, the individual spectra of
LBDS~53W091 were corrected for the effects of telluric O$_2$ absorption
using an absorption template determined from the observations of the
standard stars and scaled to the appropriate airmass. The corrected
spectra were then coadded to produce the final spectrum shown in
Figure~\ref{spectrum}.  The resultant spectrogram has an effective
exposure time of 5.5 hours.

The two red galaxies (LBDS~53W091 and 3a) have similar spectra and
similar $R - K$ colors, although the data for LBDS~53W091 are of higher
signal--to--noise ratio.  In Figure~\ref{galaxy3a} we present binned
spectra of galaxy 3a and LBDS~53W091 to illustrate their similarities;
note, in particular, the continuum discontinuity at 7400\AA.  The
spectra of the two blue galaxies (``1'' and ``3b'') are shown in
Figure~\ref{galaxy1}.  Galaxy 1, which lies 5{\arcsper}5 NW of
LBDS~53W091, shows moderately strong [\ion{O}{2}]$\lambda$3727 emission
and \ion{Mg}{2} absorption at $z=1.105$ (the \ion{Mg}{2} absorption is
affected by telluric Na~D emission).  The fainter galaxy 3b has two
weak emission lines at 5185\AA\ and 6964\AA\ which we identify as
[\ion{O}{2}]$\lambda$3727 and [\ion{O}{3}]$\lambda$5007 at $z\approx
0.4$.  Table \ref{spectroscopy} lists their emission line
identifications, fluxes and redshifts.  The spectrum shown of galaxy 1
represents 1~hour of integration on UT 95 May 27; galaxy 3b was observed
along the same long slit as LBDS~53W091 and thus represents 5.5 hours
of total integration.

\section{Results}

\subsection{Redshift Determinations}

As mentioned above, the bluer galaxies (1 and 3b) have emission line
spectra and are moderately low redshift galaxies similar to those found in
deep field surveys (\eg, Lilly \etal 1995; Cowie \etal 1995).
It is the interpretation of the two {\it red} galaxies with
absorption line spectra (LBDS~53W091 and 3a) that are the crux of this
paper, and therefore the remainder of this section describes the
determination of their redshifts.

The key to understanding the spectrum of LBDS~53W091 is the unique
``tophat''-shaped region that is observed near $\lambda \lambda 6740
- 7000$ \AA\ (see Figure~\ref{spectrum}). Inspection of the ultraviolet
spectra of F and G dwarfs obtained with the {\it Copernicus} and {\it
International Ultraviolet Explorer} ({\it IUE}) satellites clearly show
a similarly shaped feature commencing at rest wavelength $\lambda_0$
2640 \AA\ (\eg, Morton \etal 1977; Wu \etal 1991;
Figure~\ref{iuestar}). This tophat feature is caused by metal
line--blanketing on either side: in Solar type stars, the short
wavelength edge is defined largely by \ion{Fe}{2} absorption lines, and the
continuum depression on the long wavelength side is dominated by the
several weak metal lines and two strong absorption features of
\ion{Mg}{2}$\lambda 2800$\AA\ and \ion{Mg}{1}$\lambda 2852$\AA\ (in
individual spectra of G2V stars the equivalent width of the \ion{Mg}{2}
doublet is more than 25\AA; Morton \etal 1977, Fanelli \etal 1992). We
note that the observed dip in the spectrum of LBDS~53W091 at
$\lambda_{obs}\approx 6913$\AA\ coincides with a blended (and therefore
broad) telluric OH feature.  The errors in the spectrum in this
wavelength region are large and the galaxy faint, and we do not place
much weight on this particular narrow absorption feature. However, if
this feature is indeed real, it is likely due to \ion{Mg}{2}$\lambda
2800$ absorption arising in a foreground system at $z\approx 1.47$
rather than a spectral feature associated with LBDS~53W091. 

The overall shape of the observed continuum spectrum and the good match
of the continuum breaks at $\lambda_{\rm obs}\approx 6735$\AA\ and
$\approx 7500$\AA\ with the known 2640\AA\ and 2900\AA\ spectral
breaks, and the identification of the absorption feature at
$\lambda_{\rm obs} \approx 7145$\AA\ with the
\ion{Mg}{2}$\lambda$2800\AA\ doublet together suggest that the redshift
of LBDS~53W091 is $\approx 1.55$.  Cross correlation of the LBDS~53W091
spectrum in the rest wavelength range $\lambda\lambda
2100-3080$\AA\ with the spectrum of an F6V star from the Wu \etal
(1991) {\it IUE} Spectral Atlas results in a more accurate redshift of
$1.552\pm 0.002$.  The spectrum of LBDS~53W091 is also very similar to
the spectra of two nearby elliptical galaxies, M32 and NGC~3610
(Figure~\ref{m32}); this comparison adds further confidence to our
redshift determination.  Finally, we have discovered several other
galaxies with similar rest-frame spectra (Dey \etal 1997, Dickinson
\etal 1997, Stanford \etal 1997b). All of these galaxies are at
slightly lower redshifts; in several [\ion{O}{2}]$\lambda$3727 emission
and the \ion{Ca}{2} H\&K absorption lines are also detected,
reinforcing the redshift determination from the 2640\AA\ and
2900\AA\ breaks. With the exception of one object, these other galaxies
are not known to be radio sources, supporting the conclusion that these
spectral features are due to starlight.

The spectrum of the fainter red galaxy 3a (Figure~\ref{galaxy3a}) is
noisy at short wavelengths; nevertheless we can use the broad--band
colors and the observed continuum discontinuity at 7400\AA\ (which is
very similar to the rest--frame 2900\AA\ feature observed in
LBDS~53W091) to derive an estimate of its redshift.  The similarity in
all the measured broad--band colors (Table~\ref{photometry}) and the
detection of the 2900\AA\ break suggest similar redshifts for the two
galaxies, and we therefore tentatively estimate $z \approx 1.55$ for
galaxy 3a. We note that galaxy 4 also has similar colors to LBDS~53W091
(although the errors are larger), and may therefore also be at a similar
redshift.

\subsection{LBDS~53W091 as a Radio Galaxy}

LBDS~53W091 is a double--lobed FRII radio source and has a radio power
at rest--frame 1.41~GHz of $ 7.94\times 10^{33}~h_{50}^{-2}\ {\rm
erg~s^{-1}~Hz^{-1}}$. Hence, although the radio power of LBDS~53W091 is
at least 50 times less than that of the 3CR radio galaxies at similar
redshifts, it is nevertheless a fairly powerful, steep--spectrum radio source
that lies above the break in the radio galaxy luminosity function (\eg,
Fanaroff \& Riley 1974). In this subsection we discuss LBDS~53W091 in the
context of two properties of powerful radio galaxies: the alignment
effect and the uniformity of the $K$ Hubble diagram. Both of these
properties are relevant to our later discussion on the stellar content
and age of LBDS~53W091.

The intriguing aspect of the spectrum of LBDS~53W091 is that it appears
to be so similar to that of nearby normal early--type galaxies.  Most
high-redshift powerful radio galaxies have rest-frame UV spectra that
are dominated by strong line emission and non-stellar continuum
emission. The spectrum of LBDS~53W091 shows no detectable emission
lines.  [\ion{O}{2}]$\lambda$3727, usually the strongest feature in the
optical window for galaxies in the redshift range $z\approx 1.5$, is
redshifted to the very edge of our observed spectral range which is
strongly contaminated by telluric OH emission. As a result, no useful
limit can be placed on the [\ion{O}{2}] line flux.  We searched for possible
weak UV emission lines of \ion{C}{2}]$\lambda$2326 and
\ion{C}{3}]$\lambda$1909; none were found ($5\sigma$ limits are $f_{\rm
CII]},f_{\rm CIII]} \simlt 3.2 \times 10^{-18}\ {\rm
erg\ s^{-1}\ cm^{-2}}$ in the observed frame) although they would be
anticipated if an active nucleus contributed an appreciable flux of
ionizing photons at shorter wavelengths. The lack of strong emission
lines in the spectrum of LBDS~53W091 may very well be related to its
lower radio luminosity. For example, deep spectroscopy of $z\sim 1$
powerful radio sources (\eg, Stockton \etal 1996, Dey \& Spinrad 1996)
has demonstrated the presence of an underlying red stellar population
that is veiled by the strong AGN-related UV emission in the rest-frame
UV and only begins to dominate the spectrum at red rest-frame optical
wavelengths. In radio galaxies containing lower luminosity AGN, it is
therefore quite reasonable to expect that the diluting AGN continuum is
lower, and that the starlight is more easily visible, and may even
dominate the rest-frame UV spectrum.

The more powerful 3CR radio galaxies at similar redshifts ($1< z < 2$) also
show very complex, elongated rest-frame UV morphologies that tend to be
aligned with their radio axes, an indication that their morphologies
are strongly influenced by the presence of the active nucleus (McCarthy
\etal 1987, Chambers \etal 1987).  The discovery that the extended UV
continuum structures in many $z>0.7$ powerful radio galaxies are
polarized has led to the suggestion that the aligned morphologies are
caused by anisotropic radiation scattering off dust and electrons in
the ambient medium into our line of sight (\eg, di Serego Alighieri
\etal 1989).  However, it has also been suggested that the aligned UV
emission is starlight from a young stellar population formed by the
expansion of the radio source into the dense ambient medium (De Young
1981, 1989, Begelman \& Cioffi 1989).  Whichever process is
responsible, the relevant issue is whether or not one can consider the
optical light from radio galaxies as being unaffected by the presence
of the active nucleus, and therefore whether any conclusions regarding
the evolution of radio galaxies may be generally extrapolated to the
(luminous) early--type galaxy population as a whole.

If we consider galaxy 3a to be part of the LBDS~53W091 system, then it
might be argued that LBDS~53W091 exhibits the alignment effect; i.e.,
the position angle of the axis connecting the host galaxy of the radio
source to the companion galaxy 3a ($PA\approx 126\deg$) is roughly
similar to that of the radio axis ($PA\approx 130\deg$).  Since the
UV spectra of both galaxies appear to be dominated by starlight, it is
conceivable that the alignment in this system is the result of radio
source triggered star formation. However, this seems unlikely given
that both galaxies appear to be dominated by {\it old}, red
populations, whereas the radio source is fairly compact ($\approx 45$
kpc) and therefore likely young [$\approx 4.4\times 10^6 (v_{\rm
expansion}/10^4\kms)^{-1}$ yr]. It is therefore more probable that the
observed alignment is a chance coincidence. We also note that some
alignments may result from anisotropic infall along large--scale
filaments and the possible alignments observed between these filaments
and radio jet axes (e.g., West 1991).

Furthermore, the rest--frame UV spectrum argues against any significant
scattered component: the flux is roughly zero at $\lambda_{rest}\approx
1900$\AA\ and 2500\AA, and suggests that any significant scattered AGN
component would have to be at least as red as the overall galaxy
spectrum. If a reddened AGN spectrum is indeed present and
dust--scattered as is the case in most of the luminous $z\sim$ radio
galaxies, we may expect to see a wavelength--dependent image structure:
there is no evidence for this in LBDS~53W091. Finally, as discussed below, 
the 2640\AA\ and 2900\AA\ breaks are stellar absorption features and 
their amplitudes are reddening independent; the contribution of a highly 
reddened AGN component does not affect the inferences derived from these 
breaks regarding the age of the underlying stellar population.

It is well established that the $K$ Hubble diagram of powerful radio
galaxies shows remarkably little scatter ($\sigma \sim 0.5$~mag) around
a fairly linear $K$-log($z$) relation (Lilly \& Longair 1984, Lilly
1989, Eales \etal 1993). The $K-z$ sequence may be well-represented by
the predicted evolution of a passively evolving massive galaxy with a
high formation redshift.  LBDS~53W091 has a $K$ magnitude of $18.75 \pm
0.05$ and is therefore roughly 3 times brighter than an $L^*$
($M_B=-21.0$) unevolved elliptical galaxy. Note that a population
formed in an instantaneous burst at $z=5$ and evolving passively in an
\hnot=50\hnotunit, $\Omega_0=0.2$ Universe will be $\approx 1$ mag
brighter in the $K$ band at $z\approx 1.55$ than an unevolved
elliptical.  LBDS~53W091 is therefore a galaxy whose local luminosity
approximates that of an $L^*$ galaxy. Using the SED of a 3.5~Gyr old
population (from the Jimenez synthesis models; see \S~4.3) to calculate
the $K$-correction, we find rest-frame luminosities of 
$M_K\approx-27.0$ and $M_V\approx-23.9$ (for \hnot=50\hnotunit,
$\Omega_0=0.2$). 

Although LBDS~53W091 is roughly 2 times fainter ($\approx 0.75$ mag)
than the {\it mean} radio galaxy $K-z$ relation (as determined from the
3CR and 1Jy sources), it still lies within the scatter of the
Hubble diagram.  Given that the $K$-band morphology of the radio galaxy
appears undisturbed and consistent with that of an elliptical galaxy,
we conclude that the AGN contributes little light, if any, in the
observed $K$-band. 

\section{Age--Dating the UV Population in LBDS~53W091}

The similarity of the spectrum of LBDS~53W091 to the spectra of F and G
stars (Figure~\ref{iuestar}) and, in particular, to the spectra of
nearby old elliptical galaxies (Figure~\ref{m32}), suggests that this
galaxy may serve as a high-redshift benchmark in the study of the
evolution of early type galaxies. At a redshift of 1.55, an
\hnot=75\hnotunit, $\Omega_0=1$, $\Lambda=0$ universe is only 2.1~Gyr
old; hence, in principle, the age of the stellar population in
LBDS~53W091 can place strong constraints on the cosmological
parameters.

In this section we employ various methods to estimate the time elapsed
since the last major epoch of star formation in LBDS~53W091.  For the
sake of conciseness, we refer to this time as the `age' of the stellar
population. It is important to note that this age refers to the most
recent star formation episode which currently dominates the UV
spectrum, and {\it not} the first epoch of star formation in the
galaxy. Determination of the age of the UV population therefore
provides a {\it lower limit} to the age of the galaxy; the latter
should include an additional time period for the dynamical assembly of
the galaxy and the first epoch(s) of star formation (necessary to
create the observed metals and mix them into the star forming
material).

It is well known that the various extant evolutionary spectral
synthesis models result in different ages when fit to the same
optical spectra.  These differences between the models are largely due
to the differing treatments of stars in their post--main--sequence
stages (\cf Charlot, Worthey, \& Bressan 1996) as well as differing
treatments of (main--sequence) stellar spectra in the UV. We therefore
begin our analysis of the UV spectrum by deriving simple estimates of
the age which are based solely on a determination of the color of the
main--sequence turnoff population (\S~4.1 and 4.3) and comparisons to
the UV spectra of nearby elliptical galaxies (\S~4.2).  Age estimates
based on the evolutionary synthesis models are presented in \S~4.4. We
also investigate the robustness of these age estimates by applying the
same models to the UV spectrum of M32.  Since the present spectrum of
LBDS~53W091 is of insufficient signal-to-noise ratio for a detailed
comparison with the spectral synthesis models, the UV color index
$R_{UV}$ and the break amplitudes $B(2640)$ and $B(2900)$ defined below
provides a better alternative than spectral fitting for estimating the
age of the stellar population.

\subsection{The Spectral Type of the Main-Sequence Turnoff Population: 
A Semi-Empirical Approach}

The rest frame UV emission from a simple stellar population which is
older than approximately 1~Gyr is dominated by starlight from the 
main--sequence turnoff population (\eg, Charlot \& Bruzual 1991; S.\ Charlot,
personal communication). For example, Figure~\ref{4gyrstars} shows the spectrum
of a 4~Gyr-old simple stellar population (constructed using the Jimenez 
\etal (1997)
synthesis models described in \S~4.4.4) subdivided into its various
stellar evolutionary constituents, and clearly demonstrates that the
main--sequence stars completely dominate the mid-UV flux at this age. Hence, 
the determination of the effective spectral type of the
integrated UV light from the galaxy provides a fairly straightforward
measure of the mean effective temperature of the turnoff population,
and therefore an estimate of the time since the last epoch of star
formation in the galaxy. In an attempt to derive a purely empirical age
estimate for LBDS~53W091 in this section, we ignore for the present the
small contributions to the UV light from evolved stars and stars below
the main--sequence turnoff.

In order to evaluate the age of the stellar population of LBDS~53W091,
we first compared its rest frame UV spectrum ($\lambda \lambda_{\rm
rest} 1800 - 3500$) to that of F and G  stars observed by {\it IUE} (Wu
\etal 1991; kindly made available to us by Yong Li and Dave Burstein)
and to the Morton \etal (1977) spectrum of $\alpha$CMi (Procyon; F5IV
--- V) observed with {\it Copernicus}.  We constructed ``mean spectra''
of main--sequence spectral types F0V, F2-3V, F5V, F6V, F7V, F9V, G0V,
G2V, G5V and G8V by averaging together the {\it IUE} spectra of the
stars in these spectral type bins. The mean spectrum of type F6V
provided the best fit to the spectrum of LBDS~53W091 and was able to
reproduce the overall shape of the spectrum fairly accurately
(Figure~\ref{iuestar}).  This implies a color of $(B-V)\approx 0.45$
for the main--sequence turnoff population.

In order to obtain an independent estimate of the best-matching
spectral type which depends more on the details of the absorption
spectrum than on the overall shape, we define two spectral breaks,
$B(2640)$ and $B(2900)$, at the 2640\AA\ and 2900\AA\ continuum
discontinuities 
  $$B(2640) \equiv {{\bar{F_\lambda}(2645 - 2675 {\rm\AA})}
     \over {\bar{F_\lambda}(2600 - 2630 {\rm \AA})}} $$ 
  $$B(2900) \equiv {{\bar{F_\lambda}(2915 - 2945 {\rm\AA})}
     \over {\bar{F_\lambda}(2855 - 2885 {\rm \AA})}},$$
and a longer wavelength baseline UV color index
  $$R_{UV} \equiv {{\bar{F_\lambda}(3000 - 3200 {\rm\AA})}
     \over {\bar{F_\lambda}(2000 - 2200 {\rm \AA})}}$$ 
where $\bar{F_\lambda}(\lambda_1 - \lambda_2)$ is the average flux
density (in ${\rm erg\ s^{-1}\ cm^{-2}\ \AA^{-1}}$) in the wavelength
interval [$\lambda_1,\lambda_2$].  Note that our definition of the 
break amplitudes differs
slightly from that utilized in Dunlop \etal (1996).

Since the $B(2640)$ and $B(2900)$ breaks are defined over
a narrow spectral range (as indicated in Figure~\ref{m32}), they are
largely independent of reddening, and are determined primarily by the
opacities of the metal absorption lines responsible for the absorption
on their violet sides.  Table \ref{breaks} presents the measured break
amplitudes for LBDS~53W091 and compares them with those determined from
the mean F and G star spectra (see also Figures~\ref{b2640star} and
\ref{b2900star}). It is important to note that the {\it IUE} spectra
have reseaux marks that contaminate the spectral regions
$\Delta\lambda\approx 2642-2650$ and $\Delta\lambda\approx 2846-2856$
(Wu \etal 1991). These contaminate the flux at the blue edge of the
tophat feature and the \ion{Mg}{1}$\lambda$2852
absorption line. Since the tophat is roughly flat in this region, the
$B(2640)$ break determination remains unaffected.  In addition, our
definition of $B(2900)$ starts just longward of the second affected
region, and therefore this break is also fairly well determined.

Figures~\ref{b2640star} and \ref{b2900star} show the variation of the
$B(2640)$ and $B(2900)$ break amplitudes with color for main sequence stars in
the {\it IUE} spectral atlas of Wu \etal (1991).  The $B(2640)$ break
amplitude shows a significant scatter in the spectra of stars with
spectral types later than F5V, and therefore can only provide a lower
limit to the turnoff color of the UV population in LBDS~53W091 of
$(B-V)_{TO}\simgt 0.4$ (\ie, spectral types later than F5V).  The
$B(2900)$ break amplitude shows smaller scatter with spectral type or
$(B-V)$ color, and therefore provides a more robust estimate on the
color of the turnoff population of $0.55 < (B-V)_{TO} < 0.75$ (\ie, 
spectral types F9V -- G8V).  

We determined stellar age estimates as a function of metallicity and
turnoff color using the Revised Yale Isochrones (Green, Demarque and
King 1987). The results of this analysis are tabulated in Table
\ref{ryiages}. For Solar metallicities ($Z_\odot$) the bluest turnoff
color ($B-V\approx 0.45$) implies a minimum age around 2.5~Gyr.  If the
true turnoff color is $(B-V)\approx 0.6$ (as suggested by $B(2900)$),
then the corresponding turnoff age for a Solar abundance population is
$\approx 5$~Gyr.

The UV color index $R_{UV}$ also results in a consistent estimate of
the turnoff color. Figure~\ref{iueRUV} shows the variation of the UV
color index with $(B-V)$ color for the stars in the {\it IUE} spectral
atlas.  The UV color for LBDS~53W091 corresponds to a turnoff $(B-V)$
color between 0.45 and 0.55 (typical of F5V -- F9V stars) and implies a
minimum age of $\simgt 2.5$~Gyr for Solar metallicity populations.
Note that the $(B-V)$ color (and therefore age) remains roughly
constant for values of the UV color index $3.5 \simlt R_{UV} \simlt 10$
(corresponding to ages $\sim 2.5 - 5$~Gyr). This index, along with the
spectral breaks, provides a firm lower limit to the age of the
composite population.  The breaks and the overall spectrum, considered
together, imply a turnoff color of $(B-V) \simgt 0.45$ (spectral type
later than F6V), with a best fit to the break amplitudes for $(B-V)
\sim 0.6$ (spectral type G0V).  It is important to note that $R_{UV}$
is more vulnerable than the spectral breaks to reddening by dust. The
consistent estimates of the turnoff color determined from $R_{UV}$ and
the break amplitudes therefore reinforce our assumption that reddening
due to dust is minimal.

The $B(2640)$ and $B(2900)$ breaks we define above are similar to the
2609/2660 and 2828/2921 spectral breaks defined by Fanelli \etal
(1992). Studying the {\it IUE} spectra of a small sample of metal rich
and metal poor stars, Fanelli \etal found that the strengths of these
breaks are relatively insensitive to metallicity.  For LBDS~53W091, we
estimate these breaks (using the Fanelli \etal definition) to be
$0.97\pm0.24$ mag and $0.64\pm0.15$ mag respectively. These values are
typical of stars with $(B-V)\approx 0.5 - 0.6$ (of spectral type
F6V-G0V), and imply turnoff ages of $\simgt 2.5$~Gyr for populations
with $Z\le Z_\odot$.

\subsection{Comparison with Nearby Elliptical Galaxies}

As an additional empirical method to estimate the age of LBDS~53W091,
it is instructive to directly compare the spectrum of LBDS~53W091 to
the UV spectra of well--studied nearby galaxies in an attempt to
determine an age {\it relative} to the local evolved galaxy
population.  The youngest stars in a galaxy will be the bluest, and
therefore any young or intermediate--age population present will
dominate the galaxy's UV spectrum.  In Figure~\ref{m32} we plot the
normalized rest--frame UV spectral energy distribution of LBDS~53W091
along with the {\it IUE} spectrum of M32 (Burstein \etal 1988) and the
{\it HST} spectrum of NGC~3610 (Ferguson, private communication).

M32 is a nearby low luminosity galaxy which is believed
to contain an intermediate--age stellar population ($\sim 4 - 5$~Gyr old)
in addition to the very old ($\sim 10$~Gyr) stars usually present in
elliptical galaxies (\eg, Baum 1959, O'Connell 1980, Burstein \etal
1984, Rocca--Volmerange \& Guiderdoni 1987). Early studies of resolved
stars in M32 (Freedman 1992, Elston \& Silva 1992) and more recent
studies of the integrated optical and ultraviolet spectrum (Bressan
\etal 1994, Worthey 1994) are in good agreement with this conclusion,
and imply that the most recent episode of star formation in M32
occured 4 -- 5 Gyr ago. In contrast, a recent deeper imaging study with
{\it HST} by Grillmair \etal (1996) finds that the red giant branch in
M32 shows a substantial spread in color, implying that the galaxy also
exhibits a substantial range in metallicity which will affect the
interpretation of the UV light (\ie, the age of the younger
population).  Nevertheless, the youngest populations in M32 appear to have
an age of $\sim 4$~Gyr.

NGC~3610 is another well--studied nearby elliptical galaxy which shows
evidence for the presence of an intermediate--age stellar population.
NGC~3610 has an interesting morphology with twisted isophotes and a
kinematically distinct core (Scorza \& Bender 1990, Rix \& White 1992),
and shows evidence for a central stellar ring (Silva and Bothun 1997).
The galaxy colors are bluer and the nucleus shows stronger H$\beta$
absorption than similar M$_{B}$ ellipticals, though the absorption is
less than what is observed for E+A galaxies.  Furthermore, the $H - K$
color increases at the nucleus, a behavior opposite to what one expects
from dust extinction, implying an extended AGB population because AGB
stars are redder than RGB stars.  Taken together, this evidence
convincingly supports the existence of an intermediate--age population
in NGC~3610 (Silva \& Bothun 1997) similar to the more extensively
studied case of M32.  A comparison of the break amplitudes and the
$R_{UV}$ color index in NGC~3610 and M32 with those of late F and early
G stars (Table 4) strongly supports the hypothesis that an
intermediate--age population dominates the near--UV spectra in these
galaxies.

In order to compare the overall shape of the spectra, we also defined
broad spectral bins (in the ranges $2200-2400$\AA, $2650-2750$\AA, and
$2900-3100$\AA) and determined crude color indices. We note that
although M32 and NGC~3610 have composite stellar populations, the UV
light in these galaxies is very likely to be dominated by the youngest
turnoff population; the UV spectra of these galaxies therefore mimic
that of a single burst populations, and the comparison to LBDS~53W091
is therefore justified.  The spectrum of LBDS~53W091 is bluer
than the spectra of both M32 and NGC~3610, suggesting that the last
epoch of star formation in LBDS~53W091 may be slightly younger than
that in these nearby galaxies, or alternatively that LBDS~53W091 has an
additional source of UV continuum emission (cf.\ \S~4.9). Although the
UV continuum of LBDS~53W091 is bluer that that of M32 and
NGC~3610, it is important to note that within the formal errors the
amplitudes of the 2640\AA\ and 2900\AA\ breaks are roughly similar to
these systems.  We therefore estimate a minimum age of $\sim$4~Gyr for
LBDS~53W091 based upon comparison with the near--UV spectra of nearby
elliptical galaxies.

\subsection{Main--Sequence Models}

In \S~4.1 we fit the UV spectrum of the integrated light
from LBDS~53W091 with the spectrum of a single star. In this section,
we make an attempt to fit the spectrum with a composite stellar
population.  In the present approximation, we synthesize the
spectrum using a series of {\it main--sequence} stellar models. This
approach assumes that the UV emission from the galaxy is composed
entirely of starlight from main--sequence stars at and below the
main--sequence turnoff. This ignores the contribution of subgiants and
giants (the population just above the turnoff), but the impact of these
stars on the near--UV spectrum of a fairly old (${\rm \simgt 1 Gyr}$)
population should be minimal, with almost all of the $\sim \lambda 2700$\AA\ 
light arising near the main--sequence turnoff point (Charlot,
personal communication; see also Figure~\ref{4gyrstars}).

Employing the stellar atmosphere models of Kurucz (1992) and a
Miller and Scalo (1979) initial mass function (IMF), we determined the
spectral energy distribution for composite populations of different
ages by integrating the light from the total main--sequence population
(\ie, from the turnoff mass to the lower mass cutoff of the IMF). We
computed the models for three different values of the metallicity,
$Z=0.2Z_\odot$, $Z_\odot$, and $2Z_\odot$. We then compared the
continuum spectra of the resulting models and LBDS~53W091 over the
spectral range $\lambda \lambda 2000 - 3500$\AA\ (see
Figure~\ref{raulmodels}).  Since the Kurucz model atmospheres
incorporate poorly known opacities for the UV metal absorption lines
and are known to poorly reproduce some of the details of the UV spectra
of F (and later-type) stars, the hottest main--sequence star permissible
in the composite spectrum is primarily constrained by the general shape
of the spectrum and the flux at $\sim 2200$\AA.

The best fitting composite Solar metallicity main--sequence model has a
blue limiting (\ie, ``turnoff'') temperature of $T_{\rm eff} = 6900$ K
which corresponds to a stellar mass of 1.35~M$_\odot$ and a
main--sequence lifetime of 3.5~Gyr.  The stellar ages for main--sequence
stars in this mass range are robust, and are not strongly affected by
uncertainties in mass loss rates, convective overshooting, mixing
length theory, or the equation of state.  The best fitting Solar and
twice Solar metallicity main--sequence models are also able to reproduce
the 2640\AA\ break amplitude and observed $(R-K)$ color at an age of
$\simgt$ 3.5~Gyr, but do not reproduce the 2900\AA\ break or the
$(J-K)$ and $(H-K)$ colors until ages of $>$ 5~Gyr (see
Figures~\ref{breaksall} and \ref{rk}). 
The $0.2Z_\odot$ model is unable to reproduce
the breaks or the $(R-K)$ color for ages less than 6~Gyr, and the IR
colors for ages less than 13~Gyr.
We note here that the variation of the break amplitudes with age is
very similar for the main--sequence model described here and the
``full'' evolutionary model (which includes the post--main sequence
stars) described below in \S~4.4.4; this ratifies our assumption that
the breaks are dominated by starlight from the main--sequence
population of stars over the relevant range of ages.

The inconsistent ages determined from fitting the rest frame UV
spectrum (including the break amplitudes) versus those determined using
the optical and near--IR broad-band colors most likely result from the
absence of post--main--sequence stars in these models.  Another
possibility which we explore below is that LBDS~53W091 has a
composite spectrum of two stellar populations.  In populations of
ages $>$ 1~Gyr, the light at rest frame optical wavelengths (observed
near--IR) contains a significant contribution from these evolved stars,
and therefore the main--sequence models described here should only be
applied to the rest frame UV light. With this caveat in mind, the
minimum age derived from the main--sequence 
Solar metallicity models is $\approx$ 3.5~Gyr.

\subsection{Evolutionary Models}

In this section we discuss age estimates derived by fitting the
spectrum of LBDS~53W091 with the evolutionary population synthesis
models of Bruzual and Charlot (1997), Worthey (1994), Guiderdoni and
Rocca--Volmerange (1987), and our own synthesis model (Jimenez \etal
1997). (We are indebted to Drs.\ Alessandro Bressan, Stephane Charlot,
and Guy Worthey for their assistance in our model--fitting attempts,
and in the examination of the details of the models.) The Bruzual and
Charlot (1997) and the Guiderdoni and Rocca--Volmerange (1987) models
incorporate only Solar metallicity libraries (from {\it IUE} and {\it
OAO} in the UV).  The Worthey (1994) models utilize the Kurucz (1992)
theoretical stellar atmospheres as the input UV spectral library, and
therefore can be used to determine spectral sythesis ages as a function
of metallicity and thereby investigate the age--metallicity degeneracy.
It is important to note that the Kurucz model model atmospheres
incorporate poorly known opacities for the UV metal absorption lines
and therefore do not adequately reproduce some of the details of the UV
spectra of F (and later-type) stars; hence, the age of LBDS~53W091
determined from these models is primarily constrained by the general
shape of the spectrum and the flux at $\sim 2200$\AA.  We compute all
models for `instantaneous burst' star formation scenarios, \ie, star
formation lasting $\simlt 10^7\ {\rm yr}$.  The implications of this
assumption are discussed in \S~5.  We found that the spectral
discontinuities at rest wavelengths $\lambda 2640$ \AA\ and $\lambda
2900$ \AA\ as well as the UV color index (defined in \S~4.1) are useful
discriminants between the models. In Table \ref{breaks} we present the
amplitudes of these indices for LBDS~53W091, some composite F and G
stars, and the elliptical galaxies discussed in \S~4.2.  In
Figure~\ref{breaksall} we plot these breaks as a function of age for
the models discussed below.  In Figure~\ref{rk} we plot the $(R-K)$
color a function of age for these same models.

As a useful control, we also analyze M32 using the same criteria and
models.  As discussed in the previous section, M32 has an
intermediate--age stellar population ($\sim$ 4 Gyr) whose radiation
should dominate in the near--UV part of the spectrum.

\subsubsection{Bruzual--Charlot Models}

One of the most widely used evolutionary synthesis models is that of
Bruzual and Charlot (1993; see also Charlot \& Bruzual 1991, Bruzual
1983). In their present version (``BC95"; Bruzual and Charlot 1997),
these models only incorporate evolutionary tracks and spectra for stars
of Solar metallicity. These models produce very red optical--infrared
colors shortly after the initial burst of star--formation: the observed
optical--infrared color of LBDS~53W091 ($R-K=5.75$ at $z=1.55$) is
reproduced at $\approx 1.2$~Gyr (depending slightly upon the assumed
IMF) after the initial burst (see Figure~\ref{rk}).  Fitting the
overall shape of the rest frame UV spectrum results in a best-fit age
of 1.3~Gyr.  However, to also produce the spectral discontinuities of
the strengths observed in LBDS~53W091 an age in excess of 3.5~Gyr is
required.  Figure~\ref{breaksall} and \ref{rk} illustrate the
inconsistencies in population ages derived from these evolutionary
models, if a single burst is demanded for simplicity.

In Table~\ref{m32ages} we see a similar quandry when BC95 is used to
age--date M32.  The $R_{UV}$ color index yields an
extremely young ages ($\sim$1.3 Gyr) for M32, and is inconsistent
with the results discussed in \S~4.2. The break amplitudes, however,
lead to more reasonable ages of $\simgt$ 3.5~Gyr, suggesting that
greater weight should be placed on the BC95 model fits to the spectral
breaks, rather than on the fits to the overall spectrum. This procedure
then suggests a large age for LBDS~53W091: the BC95 model fits to the
spectral breaks imply ages of $\sim 6$~Gyr.  Accounting for the large
error ranges in the break amplitude measurements for LBDS~53W091, 
the BC95 models suggest a minimum age of $>2.0$~Gyr (Figure~\ref{breaksall}).

\subsubsection{Worthey Models}

Recently, models constructed by G.\ Worthey have been employed to
age--date the populations in elliptical galaxies by using indices
determined from the rest frame optical spectrum (Worthey 1994, Worthey
\etal 1996).  Dr.~Worthey has kindly computed some UV models with
metallicities of $[{\rm Fe/H}] = \pm 0.2, 0.0$ at various ages; a good
fit to the UV spectrum and the $(R-K)$ color of LBDS~53W091 occurs for
the Solar metallicity models at an age of $\sim$ 1.4~Gyr
(Figure~\ref{rk}).  However, as in the case of the Bruzual and Charlot
models, the breaks at 2640, 2900 \AA\ are not reproduced at this age.
For Solar abundance models, the 2640\AA\ and 2900\AA\ break amplitudes
are only reproduced at ages of roughly 1.5~Gyr and 4.3~Gyr
respectively.  Allowing the metallicity to vary, we find that the break
amplitudes increase more (less) rapidly for the higher (lower)
abundance models. For the three metallicities considered, no models are
capable of reproducing both the detailed spectroscopic features of
LBDS~53W091 and the broad--band colors at the same age.  In fact, these
models do not produce self--consistent age estimates for M32 either and
the 2640\AA\ break amplitude implies an exceedingly low estimate
($\sim$ 2.2 Gyr) for the age of M32 when compared with the current
literature discussed in \S~4.2.  We conclude that it is premature to
extrapolate these models, which were designed for the study of features
in the optical spectra of nearby galaxies, into the rest frame UV.

\subsubsection{Guiderdoni \& Rocca--Volmerange Models}

We also estimated the age of LBDS~53W091 using the most recent version
of the evolutionary synthesis models of Guiderdoni and
Rocca--Volmerange (1987, hereinafter G\&RV).  These models, like the
BC95 models, only incorporate a Solar metallicity stellar library, and
therefore cannot be used to investigate variations in metallicity. They
only reproduce the rest frame UV spectrum at an age of $\approx 4$~Gyr
(see Figure~\ref{paris}).  Satisfyingly, the infrared colors [$R-K
\approx 5.75, J-K\approx1.75, H-K\approx0.75$] are also reproduced at
roughly the same age, although the break amplitudes imply an even older
age ($\approx 6.5$~Gyr).  The G\&RV models are therefore roughly
self--consistent and imply a large age for LBDS~53W091.

\subsubsection{Jimenez Synthesis Models}

In order to have an independent check on the model-dependent age
estimates (cf.\ Charlot \etal 1996), we constructed our own population
synthesis code (Jimenez \etal 1997).  The code uses interior stellar
models computed using JMSTAR9 (James MacDonald, personal comm.) which
incorporates the latest OPAL opacity calculations (see Iglesias \&
Rogers 1996 and references therein); for the low temperature
atmospheres we incorporated the opacities from Alexander \& Ferguson
(1994) (Alexander, personal comm.).  Models were computed for three
values of metallicity ($0.2 Z_{\odot}$, $Z_{\odot}$ and $2
Z_{\odot}$).  Since present-day elliptical galaxies show evidence for
enhancements in $\alpha$-process elements whereas Fe-peak elements may
be under-enhanced (Worthey, Faber \& Gonzalez 1992; Weiss, Peletier \&
Matteucci 1995), we also computed tracks for $\alpha$-enhanced
metallicities to study the effects on the integrated spectra.  In
total, approximately 1000 tracks (from the contracting Hayashi track up
to the TP-AGB) were computed for stars in the mass range 0.1
$M_{\odot}$ to 120 $M_{\odot}$.  These synthesis models are similar to
the main--sequence models described in \S~4.2, but they also
incorporate the late stages of stellar evolution. We hereafter refer to
these models as the `full' models.

The code allows us to control the stellar physics that we input into
the integrated population, and it is straightforward to investigate,
for example, different mass loss laws, mixing length parameters, or
Helium abundance.  For the late stellar evolutionary stages (RGB, AGB
and HB), we used the procedure described in Jimenez \etal (1997) to
follow the evolution of stars from the base of the RGB to the TP-AGB
phase. The mass loss on the RGB and AGB was approximated using the
empirical parametrizations of Reimers (1975; see also Reimers 1977) and
Vassiliades \& Wood (1993) respectively.  This procedure allows
different scenarios for stellar evolution to be investigated quickly
and reliably.  Since the light from stars in post--main--sequence
stages of stellar evolution contribute little to the total UV emission,
the age determination using these models is insensitive to the exact
parameters chosen to calculate the late stage evolution. We were
careful not to overpopulate the post--main--sequence stages, and used
the fuel consumption theorem to compute the relative number of stars in
main sequence and  post main--sequence phases. The set of Kurucz (1992)
atmospheric models was used to computed the integrated
stellar spectra of the population.

We calculated integrated spectra for populations spanning ages from 1
to 13 Gyr, and estimated the age for LBDS~53W091 using spectral
fitting.  The lower panel of Figure~\ref{raulmodels} shows the spectrum
of LBDS~53W091 compared with synthetic spectra at three different model
ages (1, 3 and 5~Gyr). An age of 2.5~Gyr (for Solar metallicity) gives
a best fit to the overall spectrum and also matches the observed IR
colors. The UV light in the `full' models at ages $\simlt$~4~Gyr is
almost completely dominated by the main--sequence stars.  It is
therefore not surprising that these ages are in good agreement with
those derived from main--sequence models.  The effect of using the
$\alpha$-enhanced tracks was to reduce the estimated age by $\approx
0.2$~Gyr.

The model fits to the $B(2640)$ and $B(2900)$ break amplitudes yield
ages of 3.8 and 6.6~Gyr for LBDS~53W091.  The situation is similar for
M32, where the fit to the UV color index gives an age of 3.7~Gyr, while
$B(2900)$ implies an age $\sim$ 5.8~Gyr.  Comparing the model fits to the
break amplitudes and the UV color in Tables~\ref{evolages} and
\ref{m32ages}, we see that Jimenez's full models imply that LBDS~53W091
and M32 are of comparable ages.

\subsection{Summary of Age Estimation}

In Tables~\ref{evolages} and \ref{m32ages} we compare the age estimates
from our various methods.  The ``Mean Age'' column in
Table~\ref{evolages} lists the average of all the age estimates from a
given model. The different models result in a wide range of ages for
LBDS~53W091, partly due to differences in their treatments of the
post--main sequence evolutionary phases (AGB, post--AGB, horizontal
branch, etc; Charlot et al.~1996) as well as differing UV spectral
libraries ({\it IUE} versus Kurucz theoretical spectra).  As mentioned
above, the Kurucz atmospheric models incorporate poorly
known opacities for the UV metal absorption lines and inadequately
reproduce the detailed UV spectra of F (and later-type) stars. Hence,
the ages derived from most of the evolutionary synthesis models
described above are primarily constrained by the overall shape of the
UV spectrum and the spectral ``bump'' at $\lambda_{rest} \sim
2200$\AA.  We therefore place the largest weight on the age
determinations resulting from the newest models which incorporate the
most recent opacity tables (i.e., the Jimenez ``full'' models) and
those derived from the comparison of the break amplitudes measured in
LBDS~53W091 with those measured in other objects. Finally, since the
predicted $B(2900)$ break amplitude shows much less scatter among the
different models, we believe that this break provides the most reliable
age--estimate; this is endorsed by the small scatter in $B(2900)$
exhibited by the main--sequence {\it IUE} stars
(Figure~\ref{b2900star}).

Ignoring the extrema, the model fitting as a whole suggests a minimum
age of $\approx$3.5~Gyr for the population dominating the UV light from
LBDS~53W091. The $B(2900)$ break amplitude by itself suggests a lower
limit of $\approx$4~Gyr; including only the Jimenez ``full'' models and
the $B(2900)$ break amplitude results in a lower limit of
$\approx$3.4~Gyr. It is important to note that none of the age--estimates
in Table~\ref{evolages} that are based on the break amplitudes are
discrepant with a minimum age of $\sim$3.5~Gyr with the exception of
those derived from the Worthey models. The ages based on the $(R-K)$
color and the $R_{UV}$ index are also slightly lower than those
determined from the $B(2900)$ break; this is not fully understood, and
may be indicative of a mixed population (\ie, with a spread of ages;
cf., Gonz\'alez 1993), or a signature of a diluting UV component, or
simply the inadequacy of the input spectral libraries.  Whatever the
cause, it is important to note that any correction for other components
to the UV light (\eg, \S~4.9) results in even stronger break amplitudes,
and therefore larger ages. The minimum age of 3.5~Gyr resulting
from the model fitting, is therefore a strong lower limit to the age of
LBDS~53W091.

A comparison of the age--dating results for LBDS~53W091
(Table~\ref{evolages}) with a similar analysis for M32
(Table~\ref{m32ages}) suggests that the populations dominating the UV
light in these two systems have similar ages. Since the overall UV
spectrum of LBDS~53W091 is slightly bluer than that of M32, it is
likely that the $z=1.55$ galaxy is slightly younger than M32.  Given
the inconsistencies between the various models and the current
uncertain age of the UV population in M32, we will adopt the
conservative minimum age estimate of 3.5~Gyr for the remainder of this
paper.

\subsection{The IMF and Star Formation History}

The age estimates derive in the previous sections depend little on the
exact form of the IMF, so long as it is smooth and the slope and the
upper and lower mass cut-offs are reasonable.  However, the above age
estimates are all predicated on the absence of young, hot (\ie, O, B,
and A) stars in the spectrum of LBDS~53W091.  The possibility therefore
exists that the spectrum of LBDS~53W091 merely reflects an IMF devoid
of high mass stars and that the galaxy is young.  No direct evidence
exists that the galaxy contains evolved giants.  However, a truncated
IMF would be a rather contrived explanation for LBDS~53W091's spectrum,
requiring an IMF that cut off exactly at spectral type F6V to escape
the above age estimates.  With no stars $\simgt$ 1.5 M$_\odot$, the
genesis and disbursement of metals within the galaxy becomes
problematic, though the effect of these constituents are clearly
visible in the \ion{Mg}{2} 2800 absorption line as well as the spectral
breaks at 2640\AA\ and 2900\AA.  Furthermore, discussions of truncated
IMFs usually involve a suppression of the {\it low} mass stars to
escape the G--dwarf problem (\eg, Charlot \etal 1993). 
Starbursts appear to favor high--mass
stars.  We therefore find truncating the IMF a contrived explanation
for LBDS~53W091's spectrum.

The derived age estimates are also predicated on the assumption of an
instantaneous burst of star formation.  The implications of this
conservative assumption are discussed in \S~5, though the possibility
exists that the UV spectrum of LBDS~53W091 reflects multiple bursts of
star formation.  In the particular case of double--burst star formation
scenarios, the UV continuum and UV color are dominated by the youngest
stars while the break amplitudes reflect the older stars diluted by the
flatter spectrum younger population.  For example, using the BC95
models, the UV continuum slope of LBDS~53W091 is reproduced shortly
after each burst, but at these young ages the model breaks are always
overly suppressed by the hot stars, implying that no simple
double--burst scenario can satisfactorily fit all criteria
simultaneously.  This is perhaps unsurprising when one considers the
ages derived for M32 in Table~\ref{m32ages}:  it is {\it highly}
unlikely that M32 has a stellar population younger than 2~Gyr, implying
that the low ages derived from the UV continuum spectra are more
emblematic of the weaknesses of the current generation of UV
evolutionary models rather than a complicated star formation history.

\subsection{Age-Metallicity Degeneracy}

As is clear from the spectral synthesis models described in \S~4.3 and
4.4, more metal rich populations can reproduce the rest frame UV
spectrum of LBDS~53W091 at younger ages. Metal poor populations with
metallicities of $0.2Z_\odot$, on the other hand, can only reproduce the
shape of the UV spectrum and the break amplitudes at ages greater than
5~Gyr. This is the well-studied problem of the age-metallicity
degeneracy that plagues population synthesis; in the case of the nearby
elliptical galaxies, several optical spectral indices (involving
hydrogen and metal lines) have been devised to separate the effects of
age and metallicity (\eg, Worthey 1994, Gorgas \etal 1993, Jones
1996). 

Unfortunately, the optical faintness of LBDS~53W091 ($R\approx 24.5$)
precludes a direct measurement of the metallicity, especially because
the hydrogen Balmer and metal line indices commonly used to break
the age-metallicity degeneracy are redshifted into the near--IR for
redshifts $z>1.2$.  It is possible that future efforts with infrared
spectrographs and adaptive optics on large telescopes will permit a
measurement of abundances in the integrated light using features that
are well-studied in local elliptical galaxies. For the present, the
age estimates of LBDS~53W091 remain degenerate with
metallicity.

However, if we assume that LBDS~53W091 is a progenitor of a present-day
elliptical galaxy and that no active star formation has taken place
over the last $\sim$ 11 Gyr (lookback time to $z=1.55$ for \hnot=50,
$\Omega_0=0.2$), then the mean metallicity of LBDS~53W091 should be
very similar to that found for local $\simgt L^*$ elliptical galaxies.
Nearby elliptical galaxies with luminosities larger than $L^*$
generally have spatially integrated metallicities (determined from the
integrated spectrum) that are approximately Solar (Worthey \etal 1984;
Kormendy \& Djorgovski 1989; Worthey, Faber \& Gonz\'alez 1992; Sadler
1992).  Within the effective radius $R_e$, the luminosity--averaged
metallicity of nearby luminous ellipticals is roughly Solar.  For
example, the Mg index of luminous ellipticals is $\approx$ 0.30 in the
nuclear regions (\ie, super--Solar metallicity of $[{\rm Mg/H}] = +0.2$
in the core), whereas it decreases to $\approx$ 0.22 at radii near
$R_e$ (Buzzoni 1996), implying sub--Solar metallicities, $[{\rm Mg/H}]
\sim -0.3~{\rm to}~-0.4.$ Gonzalez \& Gorgas (1996) present Mg index
profiles for several giant ellipticals, again suggesting a similar mean
Mg index of $\sim 0.22$ for radii of 0.5 $R_e$ to $R_e.$  Arimoto
(1996) concludes that a mean metallicity of Solar is appropriate for
nearby ellipticals based on measurements of the abundance in the hot
corona and old stars.

The effective radius, $R_e,$ of an $L^*$ galaxy at $z = 1.55$ is
$\approx 0\farcs7$ (\eg, Dickinson 1995).  Hence, our
1\arcsec\ spectroscopic slit samples the galaxian profile to almost
$R_e.$ It is therefore reasonable to assume that the UV population in
LBDS~53W091 has a metallicity which is approximately Solar within the
aperture of our observations.  Even for twice Solar metallicity, the
break amplitudes imply an age in excess of 3~Gyr for LBDS~53W091 (see
Dunlop \etal 1996, Figure~2c).

\subsection{Dust Reddening and the Age Limit}

Thus far, we have avoided the conclusion that dust reddening of a young
stellar population is responsible for the red color of LBDS~53W091.
The most important objection to strong dust reddening is that stellar
populations younger than $\approx 2-3$~Gyr fail to reproduce the
spectral features observed in the rest frame UV spectrum. In addition,
the $B(2640)$ and $B(2900)$ spectral breaks used in the preceding
discussion are defined over short, adjacent wavelength intervals of the
UV spectrum, and are therefore virtually reddening independent. We
noted earlier that in most cases, the $B(2640)$ and $B(2900)$ continuum
breaks suggested an age similar to that implied by the broader baseline
$R_{UV}$ color index, suggesting that the reddening is negligible.  As
a test, we dereddened the spectrum using an LMC extinction law and an
$E(B-V)=0.1$ (\ie, $A_B\approx 0.4$). The best fitting Bruzual and
Charlot model to this dereddened spectrum has an age of $\approx
1.3$~Gyr but provides a poor fit to the break amplitudes and the
optical-IR colors.

For the remainder of this paper we consider $3.5$~Gyr a minimum for galaxy
LBDS~53W091. A similar minimum age is likely to apply to galaxy 3a, by
virtue of its spatial proximity to LBDS~53W091 and its similar spectral
energy distribution.  The implications of finding two very red galaxies
in close proximity and at similar redshifts is discussed in \S~5.

\subsection{Other Contributions to the UV Light}

\subsubsection{Active Galactic Nucleus}

The AGN may contribute to the ultraviolet continuum emission in
LBDS~53W091, either directly or by dust and electron scattering as it
typically does in more powerful radio galaxies (\eg, di Serego
Alighieri \etal 1989; Cimatti \etal 1993; Jannuzi \etal 1996; Dey \&
Spinrad 1996).  The AGN continuum would tend to dilute the spectral
features and the break amplitudes, and hence accounting for this
component in the spectrum would make the intrinsic break amplitudes
even larger and imply an even older age.  Since the breaks at
2640\AA\ and 2900\AA\ are already strong, are comparable to those in
individual stars, and fail to be reproduced by most of the population
synthesis models, it is unlikely that the AGN contribution is
significant at UV wavelengths. A second indication that the AGN
contribution is likely to be very minimal is the apparent lack of
strong emission lines in the UV spectrum: the limits on the
\ion{C}{2}], \ion{C}{3}]  and \ion{Mg}{2} emission lines are roughly 10
times fainter than that observed in powerful (3CR) radio galaxies at
similar redshifts (McCarthy 1993).

We can make a rough estimate of the UV contribution from an AGN in 
LBDS~53W091 by comparing it with the powerful 3CR radio galaxies. The 
radio power at 1.4~GHz of LBDS~53W091 is approximately 50 times smaller 
than that of a typical 3CR radio galaxy at $z\simgt 1.5$. If the 
UV luminosity of the AGN is also reduced by this factor, the 
observed $R$ band magnitude (rest frame $\sim 2700$~\AA) of LBDS~53W091 
would be $\approx 26$ mag, or approximately one third of the observed
near--UV flux.  This contribution, if present, would only dilute
the break amplitudes, implying an even greater age for LBDS~53W091. Since
the present age estimate already provides stringent constraints to the 
cosmological parameters, it is unlikely that the diluting AGN contribution 
is significant. 

\subsubsection{Blue Stragglers}

In old Galactic star clusters, blue stragglers (thought to be hot
binary stars or stellar merger remnants) can contribute significantly
to the total short wavelength UV flux from the cluster.  These stars
are brighter, and often considerably bluer than the stars near the 
main--sequence turnoff in cluster color--magnitude diagrams.  Most
importantly, they are not represented in any of the contemporary
theoretical isochrones used by extant spectral synthesis models. Thus,
if the integrated spectra of galaxies are similar to those of old
Galactic clusters and contain a contribution from a blue straggler
population, the present synthesis models will underestimate their age.

To examine the situation quantitatively, we utilized color--magnitude
arrays (Milone \& Latham 1994) and a luminosity function from M67
(Montgomery \etal 1993, Fan \etal 1996) for the clusters listed in Table
\ref{bstrags}.  We crudely estimated the blue straggler contribution to
the integrated UV light from clusters ($\lambda_{\rm rest}\sim
2600$\AA) under the assumption that the blue stragglers have UV spectra
resembling their main--sequence ($B-V$) analogs and assuming that the
mass function determined for M67 (Montgomery \etal 1993) is applicable
to all clusters.  For M67 itself, our most robust blue straggler case,
these stars contribute approximately one half of the total light at
2600\AA.  At the other extreme, the solitary bright blue straggler star
in NGC~752 makes up only 20\% of the integrated UV flux from the
cluster.  The other clusters listed in Table \ref{bstrags} lie roughly
between these extremes.

Hence, if the stellar content of LBDS~53W091 is similar to that of the
Galactic clusters, blue stragglers {\it may} be responsible for as much as
$\sim 20\% - 50\%$ of the UV flux. Accounting for this contribution
will, as in the case of the AGN, increase the age of the turnoff
population.  Our reliance on the isochrone models described in \S~4.3
for estimating the age of LBDS~53W091 spectrum is undoubtedly naive;
however, most of the substantive uncertainties point toward our mean
age of 3.5~Gyr (or {\it any} age determined using these models)
being a lower bound.

\section{On the Formation History of LBDS~53W091 and Cosmological Implications}

The 3.5~Gyr minimum age we deduce in \S~4 is almost certainly a
significant under--estimate of the true age of LBDS~53W091.  First,
this age assumes that the fairly large elliptical galaxy was formed in
an instantaneous stellar burst after which star formation completely
ceased.  More realistically, the initial episode of star formation is
likely to last at least one dynamical collapse time, $\gtrsim 2 \times
10^8$ yrs.  If one assumes an extended episode of star formation, the
derived total age increases in direct agreement with the assumed
duration of the star formation burst; the ages derived in \S~4 are
actually the time elapsed since the cessation of star formation in the
galaxy, since the UV spectrum for old populations is dominated by stars
at the main--sequence turnoff.  Second, it is rather unlikely that the
turnoff population that dominates the UV starlight is a result of the
{\it first} episode of star formation. Because the metallicity of the
population is almost certainly non-primordial, the gas from which the present
UV--dominant population was formed must have been enriched by previous
episodes of star formation. The duration of these previous star forming
episodes, and the time between the earlier episodes and the present
one, must also be added to the age of the galaxy. Future spectroscopic
observations of LBDS~53W091 in the near-- and mid--IR may allow us to
determine its giant content and thereby constrain the contribution from
previous bursts to the integrated spectrum.  Therefore, in accounting
for the original dynamical collapse time of the galaxy, and multiple,
non--instantaneous episodes of star formation, the adopted `age' of
3.5~Gyr is found to provide a conservative lower bound to the true
age of the galaxy.

Independent of cosmology, the discovery of a high redshift galaxy with
a spectrum nearly identical to that of nearby, old elliptical galaxies
has the profound implication that the epoch of formation of these early
type systems must be at very high redshifts ($z\ge5$). If the other red
galaxies which lie nearby (in projection) are indeed physically
associated with LBDS~53W091, they raise the additional problem of an
early epoch for structure formation. 

An old galaxy at $z=1.552$ can impose strong constraints on the
time--scale for cosmology, the epoch of the last burst of
star--formation and, perhaps, the epoch of galaxy assembly.  We
consider first cosmologies without a cosmological constant ($\Lambda =
0$).  Figure~\ref{cosmoplot}a shows the parameter space of the $H_0 -
\Omega_0$ plane allowed by the existence of a 3.5~Gyr old galaxy at a
redshift $z=1.552$ (the hatched region is excluded).  Recent
measurements of $H_0$ (Kennicutt \etal 1995; Sandage \etal 1996)  imply
values between 50 and 80 km s$^{-1}$ Mpc$^{-1}$.
Figure~\ref{cosmoplot} simply re--illustrates the familiar time--scale
problem resulting in studies of the ages of Galactic globular
clusters.  In the present case, the age problem is referred to a time
when the Universe was less than 30\% of its present age, and the
uncertainties are largely independent of those encountered in the
globular cluster studies. The existence of LBDS~53W091 permits only low
Hubble constants and/or low cosmic densities; in particular, an $\Omega
= 1$ Universe requires $\hnot \simlt 45\hnotunit$.  With $H_0 = 50$ km
s$^{-1}$ Mpc$^{-1}$, a Universe with $\Omega_0 \lesssim 0.2$ is
acceptable; for this cosmology we derive a formation redshift for
LBDS~53W091 age of $z_f \ge 5.$

A possible solution to the age paradox is to invoke a
non--zero cosmological constant.  Figure~\ref{cosmoplot}b illustrates
the constraints on the $\hnot - \Omega_\Lambda$ parameter space for a
flat ($\Omega_{\rm total} = \Omega_0 + \Omega_\Lambda = 1$) universe
imposed by a 3.5~Gyr old galaxy at a redshift of $z = 1.552.$  {\it
HST} counts of ellipticals down to $I \approx 24.5 (B \approx 26.5)$
imply $\Omega_\Lambda \leq 0.8,$ with a likely range of $\Omega_\Lambda
\lesssim 0.5$ (Driver \etal 1996).  {\it COBE} measurements of the
cosmic microwave background imply a similar upper limit,
$\Omega_\Lambda \leq 0.5$ for $\hnot = 70$ km s$^{-1}$ Mpc$^{-1}$ (White
\& Bunn 1995), as do analyses of gravitational lens statistics
($\Omega_\Lambda < 0.66$ at the 95\% confidence level, Kochanek 1996)
and high--redshift supernovae ($\Omega_\Lambda < 0.51$ at the 95 \%
confidence level, Perlmutter \etal 1996).  LBDS~53W091 implies a {\it
lower} limit to $\Omega_\Lambda$ (for $\Omega_{\rm total} = 1$):  for
$\hnot > 50$ km s$^{-1}$ Mpc$^{-1},$ we find $\Omega_\Lambda \simgt
0.15,$ while for $\hnot > 70$ km s$^{-1}$ Mpc$^{-1}, \Omega_\Lambda
\simgt 0.5.$  For certain values of the cosmological parameters,
LBDS~53W091 thus provides tighter (and independent) constraints than
the well--known globular cluster age limits.

If the old, red, and ``dead'' elliptical galaxies that we now observe
at intermediate redshifts ($z\simlt 1$) really did form this early, and
if their initial starburst phase had a short duration, some luminous
galaxies near $z = 6$ should eventually be observable in the near--IR
domain, and should be identifiable by their Lyman limit cutoff in the
optical part of the spectrum.  If, however, the typical formation
redshift is much larger (\eg, $z_f \ge 10$), these elusive objects
await discovery by NICMOS on the {\it HST}.

\section{Conclusions}

We have observed the weak radio source LBDS~53W091, associated with a
very faint red galaxy ($R \approx 24.5, R - K \approx 5.8$).  Deep exposures
with the W.M. Keck telescope reveal a spectrum devoid of strong
emission lines, and dominated by starlight from a red stellar
population.  Based on the 2640\AA\ and 2900\AA\ spectral breaks, we
determine the absorption line redshift of the galaxy to be $z=1.552 \pm
0.002$.  The rest-frame UV spectrum, generally dominated by the
main-sequence turnoff population in intermediate--age coeval
populations, is similar to that of late F stars. The best--fit turnoff
spectral type of F6V suggests a strict lower limit of $\sim$~2.5~Gyr for the 
age of LBDS~53W091, implying that it is the oldest galaxy yet discovered at $z
\simgt 1.$ It is important to note that the amplitudes of the UV 
continuum spectral breaks at 2640\AA\ and 2900\AA, as well as the
broader baseline UV color index are all consistent with a main sequence
turnoff color of $(B-V)\approx 0.5$ (\ie, a spectral type of F6V). Since
the UV color index is more easily affected by dust than the spectral
breaks, the consistent turnoff color estimates strongly suggest that
the dust reddening in LBDS~53W091 is minimal.  The rest-frame UV
spectrum of LBDS~53W091 is very similar to (albeit slightly bluer) than
that of the well--studied nearby ellipticals M32 and NGC~3610. Since
the UV light in these nearby systems is dominated by an
intermediate--age stellar population ($\sim$~4--5~Gyr) in addition to
the old population typical of ellipticals, the population dominating
the UV light in LBDS~53W091 is likely to be of comparable age.

We have also estimated the age of LBDS~53W091 (\ie, the time elapsed
since the last major epoch of star formation) using a variety of
spectral synthesis models.  Using the synthesized spectra of composite
{\it main--sequence} stellar populations of varying metallicity, we
find a best fit age of 3.5 Gyr for Solar metallicity.  We also fit the
spectrum using the current evolutionary population synthesis models of
Bruzual and Charlot (1997), Jimenez \etal (1997), Worthey (1994), and
Guiderdoni and Rocca--Volmerange (1987).  We find that the different
models do not result in self--consistent ages for either LBDS~53W091 or
the nearby, well--studied elliptical M32. These inconsistencies are
likely due to differing treatments of stars in their evolved stages, as
well their reliance on differing UV stellar spectral libraries and the
uncertainties in the metallicities.  The most robust self--consistent
age estimates result from model (and single star) fits to the
2900\AA\ break amplitude, and from the models which incorporate the
newest opacity tables.  We conservatively combine the various age
estimates and derive a {\it minimum} age of 3.5~Gyr for LBDS~53W091.
Finding such an old galaxy at these large lookback times has important
cosmological consequences.  In particular, this result effectively
rules out $\hnot \simgt 45$ km s$^{-1}$ Mpc$^{-1}$ for $\Omega = 1.$

\acknowledgements

We are grateful to Mark Dickinson, Wayne Wack, Terry Stickel, Randy
Campbell and Tom Bida for their invaluable help on our Keck observing
runs. We are also very grateful to Alessandro Bressan, Stephane
Charlot, Ben Dorman and Guy Worthey for their generous help and advice
on the various stellar spectral synthesis models presented in this
paper.  We thank Yong Li and Dave Burstein for providing us with the
digitized version of the {\it IUE} stellar spectral atlas, and Dave
Burstein, Harry Ferguson, and Mike Eracleous for providing us with the
UV spectra of M32 and NGC~3610.  We thank John Davies for carrying out
the UKIRT service observations, Dave Silva for useful discussions
regarding nearby ellipticals and Ata Sarajedini for providing us with
the most recent version of the Revised Yale Isochrones. Finally, we
thank the referee Jim Schombert for an extremely prompt and useful
referee report.  The W.\ M.\ Keck Telescope is a scientific partnership
between the University of California and the California Institute of
Technology, made possible by a generous gift of the W.\ M.\ Keck
Foundation. The United Kingdom Infrared Telescope is operated by the
Royal Observatories on behalf of the UK Particle Physics and Astronomy
Research Council. The National Optical Astronomy Observatories are
operated by the Association of Universities for Research in Astronomy
under Cooperative Agreement with the National Science Foundation.  This
work was supported by the US National Science Foundation (grant \#
AST-9225133 to HS and AST-8821016 to RAW), by an Alfred P.\ Sloan
Fellowship to RAW and an EC Fellowship to RJ.

\vfill\eject

\begin{deluxetable}{lllcl}
\tablewidth{0pt}
\tablecaption{Radio Data\tablenotemark{\dag}}
\tablehead{
\colhead{Component} & \colhead{$RA_{1950}$} & \colhead{$DEC_{1950}$} &
\colhead{$\nu\ ({\rm GHz})$} & \colhead{$F_\nu\ ({\rm mJy})$}}
\startdata
Total & ${17^h 21^m 17{\secper}81\pm 0{\secper}01}$ & ${+50^\circ 08^\prime 47{\arcsper}6 \pm 0{\arcsper}1}$ & 1.565 & $23.0 \pm 1.7$ \nl
             &    &    &  4.860 & $6.5 \pm 0.4$\nl
SE Lobe & $17^h 21^m 17{\secper}98\pm 0{\secper}01$ & $+50^\circ 08^\prime 46{\arcsper}18 \pm 0{\arcsper}05$ & 1.565 & $11.5 \pm 1.3$ \nl
                  &    &    & 4.860 & $3.37 \pm 0.23$\nl
NW Lobe & $17^h 21^m 17{\secper}64\pm 0{\secper}01$ & $+50^\circ 08^\prime 49{\arcsper}00 \pm 0{\arcsper}07$ & 1.565 & $10.7 \pm 1.3$ \nl
                  &    &    & 4.860 & $2.25 \pm 0.29$\nl
\enddata
\tablenotetext{\dag}{Data in this table are derived from the 1995 VLA observations described in the text.}
\label{radiodata}
\end{deluxetable}

\begin{deluxetable}{llclll}
\tablewidth{0pt}
\tablecaption{Photometry in the LBDS~53W091 Field.}
\tablehead{
\colhead{} & \colhead{Galaxy 1} & \colhead{LBDS~53W091} &
\colhead{Galaxy 3a} & \colhead{Galaxy 3b} & \colhead{Galaxy 4}}
\startdata
$R$ &$ 23.9\pm0.1 $&$ 24.5\pm0.2 $&$ 24.9\pm 0.2 $&$ 25.1\pm0.3 $&$ 25.5\pm 0.3 $\nl
$J$ &$ 22.1\pm0.5 $&$ 20.5\pm0.1 $&$ 20.5\pm 0.1 $&$ 22.2\pm0.5 $&$ 20.6\pm 0.2 $\nl
$H$ &$ 21.5\pm0.4 $&$ 19.5\pm0.1 $&$ 19.5\pm 0.1 $&$ 21.5\pm0.4 $&$ 20.0\pm 0.1 $\nl
$K$ &$ 19.8\pm0.3 $&$ 18.7\pm0.1 $&$ 18.9\pm 0.2 $&$ 20.1\pm0.5 $&$ 19.0\pm 0.3 $\nl
\enddata
\tablecomments{All magnitudes are measured in a 4\arcsec\ diameter aperture.}
\label{photometry}
\end{deluxetable}
 
\begin{deluxetable}{lllcl}
\tablewidth{0pt}
\tablecaption{Line Identifications in the Blue Galaxies.}
\tablehead{
\colhead{Source}                   & \colhead{$\lambda_{\rm obs}$}		&
\colhead{Line ID}		   & \colhead{Flux}		&
\colhead{$z$}	\nl
\colhead{} & \colhead{\AA} & \colhead{} & \colhead{(${\rm 10^{-17}~erg\ cm^{-2}\ s^{-1}}$)} &
\colhead{}}
\startdata
Galaxy 1	& 5897:		& \ion{Mg}{2}   & abs. 	  & 1.105 	\nl
		& 7846.5	& [\ion{O}{2}]	& 7.0 	  & 1.105	\nl
		&		&	&	& $\bar{z}=1.105$ 	\nl
		&		&	&	&			\nl
Galaxy 3b	& 5185		& [\ion{O}{2}]	& 0.5     & 0.391	\nl
		& 6964		& [\ion{O}{3}]	& 0.4     & 0.391	\nl
		&		&	&	& $\bar{z}=0.391$ 	\nl
\enddata
\label{spectroscopy}
\end{deluxetable}

\begin{deluxetable}{lccccl}
\tablewidth{0pt}
\tablecaption{Break Amplitudes.}
\tablehead{
\colhead{Object} & \colhead{$B(2640)$} & \colhead{$B(2900)$} & 
\colhead{$R_{UV}$} & \colhead{$B-V$} & \colhead{Notes}}
\startdata
F0V         & 1.69  & 1.24 &  1.90 &  0.31   & {\it IUE} \nl
F2-3V       & 1.69  & 1.19 &  2.27 &  0.36   & {\it IUE} \nl
F5V         & 2.04  & 1.23 &  3.86 &  0.43   & {\it IUE} \nl
F6V         & 2.42  & 1.33 &  5.46 &  0.45   & {\it IUE} \nl
F7V         & 2.38  & 1.34 &  6.38 &  0.48   & {\it IUE} \nl
F9V         & 2.42  & 1.47 &  8.50 &  0.57   & {\it IUE} \nl
G0V         & 2.73  & 1.59 & 15.88 &  0.59   & {\it IUE} \nl
G2V         & 2.63  & 1.70 & 24.59 &  0.63   & {\it IUE} \nl
G5V         & 2.51  & 1.97 & 35.70 &  0.66   & {\it IUE} \nl
G8V         & 2.61  & 2.13 & 34.32 &  0.74   & {\it IUE} \nl
            &       &      &       &         & \nl
M32         & 2.02  & 1.59 & 5.49  &       & {\it IUE} \nl
NGC~3610    & 2.02  & 1.62 & 19.08 &       & {\it HST} \nl
            &       &      &       &         &  \nl
LBDS~53W091 & 2.27$\pm$0.35 & 1.70$\pm$0.26& 3.94$\pm$0.52 &  & Keck\nl
\enddata
\label{breaks}
\end{deluxetable}

\begin{deluxetable}{lcc}
\tablewidth{0pt}
\tablecaption{Yale Isochrone Ages ($Y=0.2$)}
\tablehead{
\colhead{$Z$} & \colhead{Age (Gyr)} & \colhead{Age (Gyr)} \nl
\colhead{} & \colhead{$B-V=0.45$} & \colhead{$B-V=0.60$}
}
\startdata
0.004  & 7.4  & 20.3 \nl
0.01   & 4.4  & 10.4 \nl
0.02\tablenotemark{\dag}& 2.5  & 5.1  \nl
0.04   & 1.8  & 3.5 \nl
0.1    & 1.5  & 2.6 \nl
\enddata
\label{ryiages}
\tablenotetext{\dag}{Interpolated from neighbouring metallicities.}
\tablecomments{The metallicity of the Sun is $Z_\odot \equiv 0.02$ by
definition for the Revised Yale Isochrones.}
\end{deluxetable}

\begin{deluxetable}{lccccc}
\tablewidth{0pt}
\tablecaption{Evolutionary Model Ages: LBDS~53W091}
\tablehead{
\colhead{Model} & \colhead{$B(2640)$} &
\colhead{$B(2900)$} & \colhead{$R_{UV}$} & \colhead{$R-K$} & 
\colhead{Mean Age}} 
\startdata
{\it IUE} & $\simgt$ 2.5  & 5.1 & $\simgt$ 2.5  & \nodata & $\simgt$ 3.4  \nl
\medskip
Jimenez--MS   & $4.2^{+1.0}_{-1.0}$ &
	  $6.5^{+2.4}_{-1.6}$ & $3.3^{+0.2}_{-0.3}$ &
	  $4.6^{+0.4}_{-0.2}$ & 4.7 \nl
\medskip
BC95	& $6.5^{+4.5}_{-4.5}$ & 
	  $6.0^{}_{-3.5}$        & $1.3^{+0.1}_{-0.1}$ &
	  $1.2^{+0.2}_{-0.1}$ & 3.8 \nl
\medskip
Jimenez--full & $3.8^{+1.2}_{-1.1}$ &
	  $6.6^{+3.1}_{-2.1}$ & $2.8^{+0.3}_{-0.3}$ &
	  $2.5^{+0.4}_{-0.2}$ & 3.9 \nl
\medskip
Worthey	 & $1.5^{+0.6}_{-0.4}$ & 
	  $4.3^{+2.7}_{-1.3}$ & $1.6^{+0.1}_{-0.2}$ &
	  $1.2^{+0.2}_{-0.1}$ & 2.2 \nl
\enddata
\label{evolages}
\tablecomments{Age ranges estimated from 1$\sigma$ errors of LBDS measurements.}
\end{deluxetable}

\begin{deluxetable}{lcccc}
\tablewidth{0pt}
\tablecaption{Evolutionary Model Ages: M32}
\tablehead{
\colhead{Model} &  \colhead{$B(2640)$} &
\colhead{$B(2900)$} & \colhead{$R_{UV}$} & \colhead{Mean Age} } 
\startdata
{\it IUE}	& $\simgt$2.5 & 5.1 & $\simgt$2.5 & $\simgt$3.4 \nl
Jimenez--MS   	&  3.5 & 5.8 & 4.1 & 4.5 \nl
BC95		&  3.5 & 4.0 & 1.3 & 2.9 \nl 
Jimenez--full 	&  3.0 & 5.8 & 3.7 & 4.2 \nl
Worthey		&  1.3 & 3.2 & 2.0 & 2.2 \nl
\enddata
\label{m32ages}
\end{deluxetable}

\begin{deluxetable}{lccl}
\tablewidth{0pt}
\tablecaption{Confirmed Blue Stragglers in Open Clusters.} 
\tablehead{
\colhead{Cluster} & \colhead{Age (Gyr)} &
\colhead{Blue Stragglers} & \colhead{References}}
\startdata
NGC~6939	& 1.6	& $\geq 1$ 	& a \nl
NGC~2360	& 1.9	& $\geq 1$ 	& a \nl
NGC~7789	& 2	& $\geq 7$  	& a \nl
NGC~752		& 2.4	& 1 	    	& a \nl
NGC~2420	& 4	& $\geq 2$ 	& a \nl
NGC~2682 (M67)	& 5	& $\geq 10$ 	& a,b \nl
NGC~188		& 6	& $\sim 11$	& c \nl
\enddata
\tablerefs{a: Milone \& Latham 1994; b: Montgomery \etal 1993; c: Dinescu \etal
1996}
\label{bstrags}
\end{deluxetable}



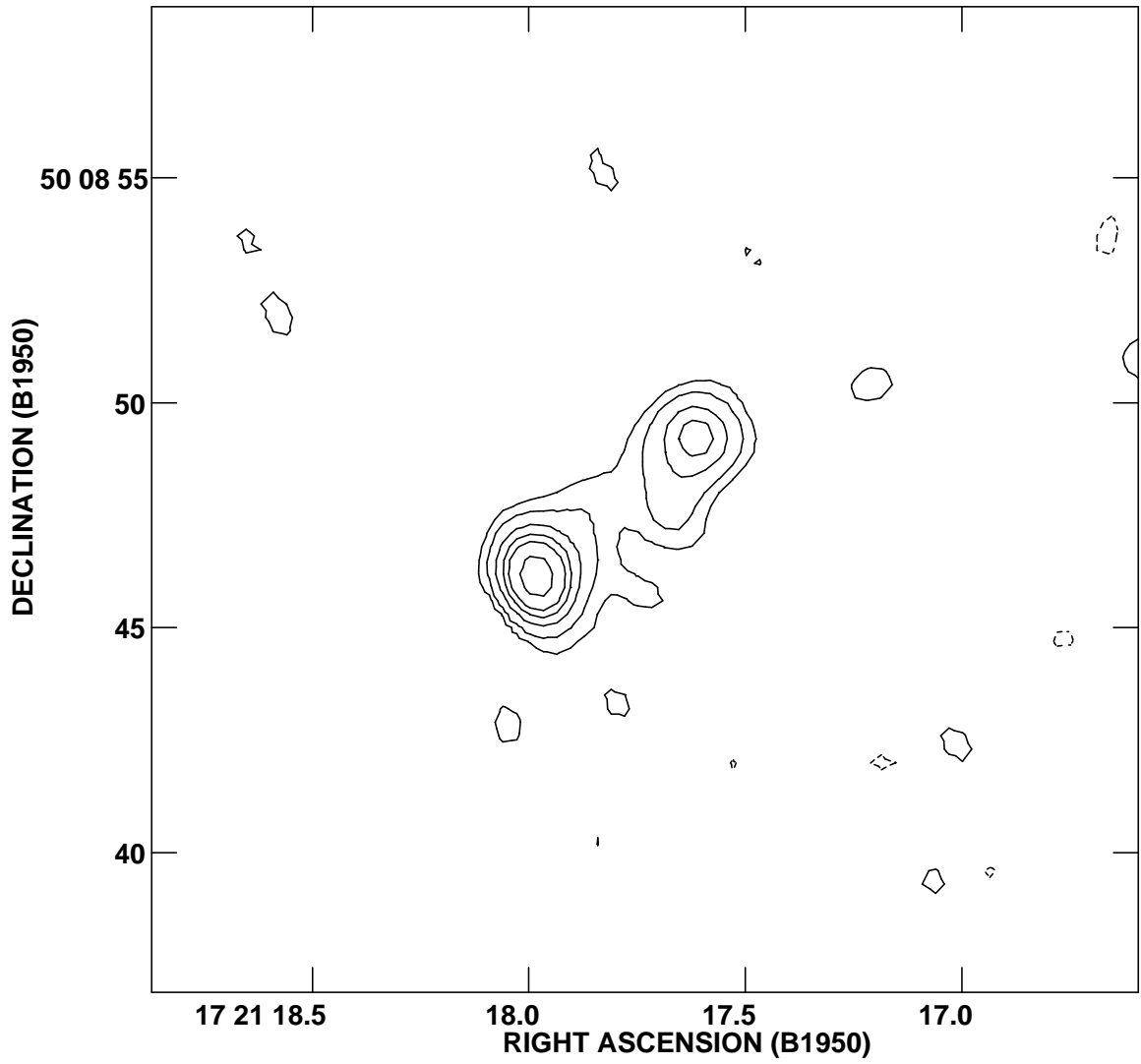
\figcaption{VLA A-Array 4.86~GHz map of the radio source LBDS~53W091. The 
noise in the map is $\sigma = 52\mu$Jy/beam, and the contours shown are 
drawn at ($-$3,3,6,12,18,24,36)$\sigma$. \label{vla}}


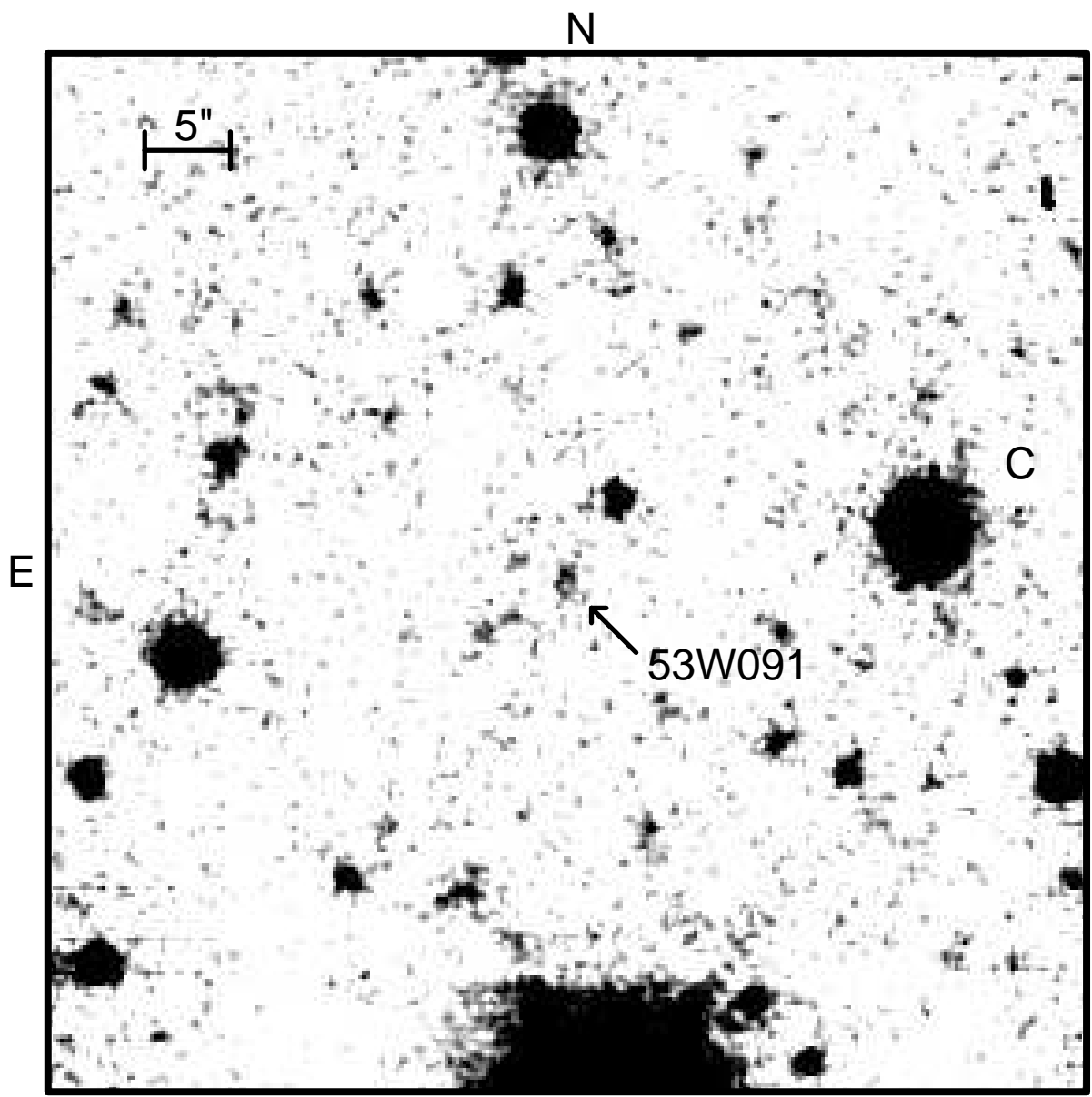
\figcaption{Keck $R$-band of the field of LBDS~53W091. The frame is
$1'$ on a side; north is to the top and east is to the left. The scale
bar shown at top left corresponds to $\approx 55.7$~kpc at $z=1.552$.
The optical counterpart of the radio source is at $\alpha_{1950} =
17^h21^m17{\secper}78, \delta_{1950} = 50\deg 08^\prime 47{\arcsper}3$,
and the offset from galaxy C to LBDS~53W091 is $\Delta\alpha=20{\arcsper}5$
(east), $\Delta\delta=-2{\arcsper}8$ (south).\label{chart}}


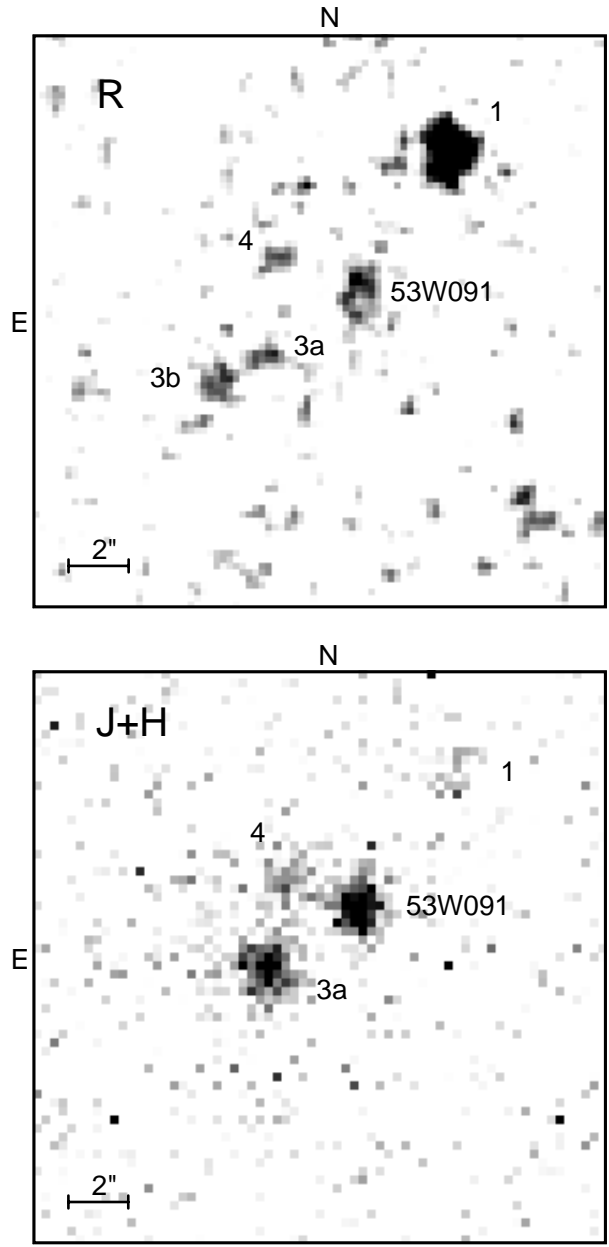
\figcaption{(a) Detail of the Keck $R$-band image of LBDS~53W091. (b)
Sum of the UKIRT $J$ and $H$ band images. Both frames are 19\arcsec\ on
a side, and north is to the top and east is to the left.  The host
galaxy of the radio source is labelled 53W091.  The blue objects 1 and
3b are foreground emission line galaxies.  Object 3a and 4 have similar
optical--IR colors to LBDS~53W091 and are likely to be at the same
redshift. \label{images}}


\figcaption{False color image of the field of LBDS~53W091 constructed
using the images in the $R$--band (blue), $J$--band (green), and
$H$--band (red) of the field of LBDS~53W091. Note that the host galaxy
of the radio source and the two objects nearest it have roughly the
same color, and may be all at a common redshift.  \label{colorpic}}


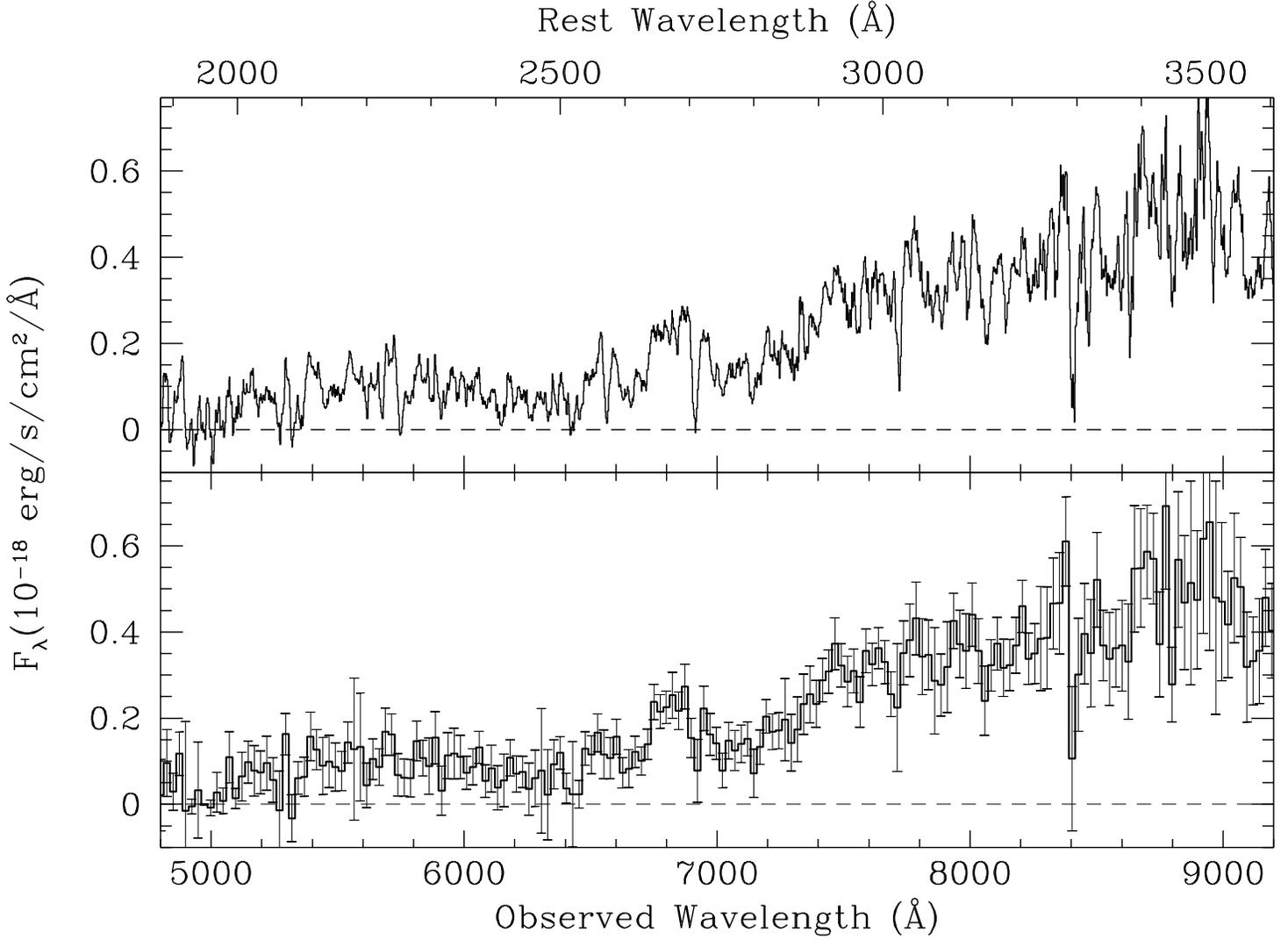
\figcaption{The 5.5 hour Keck LRIS spectrum of the host galaxy of
LBDS~53W091 plotted in the observers' frame.  The upper
panel shows the coadded spectrum smoothed using a boxcar filter of
width 9 pixels. The lower panel shows the formal 1$\sigma$ error bars
on the spectrum (averaged in 10-pixel bins). The rest wavelength is
indicated along the upper abscissa for a redshift of $z=1.552.$ The
long wavelengths suffer increased noise from atmospheric OH emission
lines.  The spectrum has been corrected for telluric O$_2$ absorption
in the A-- and B--bands. \label{spectrum}}


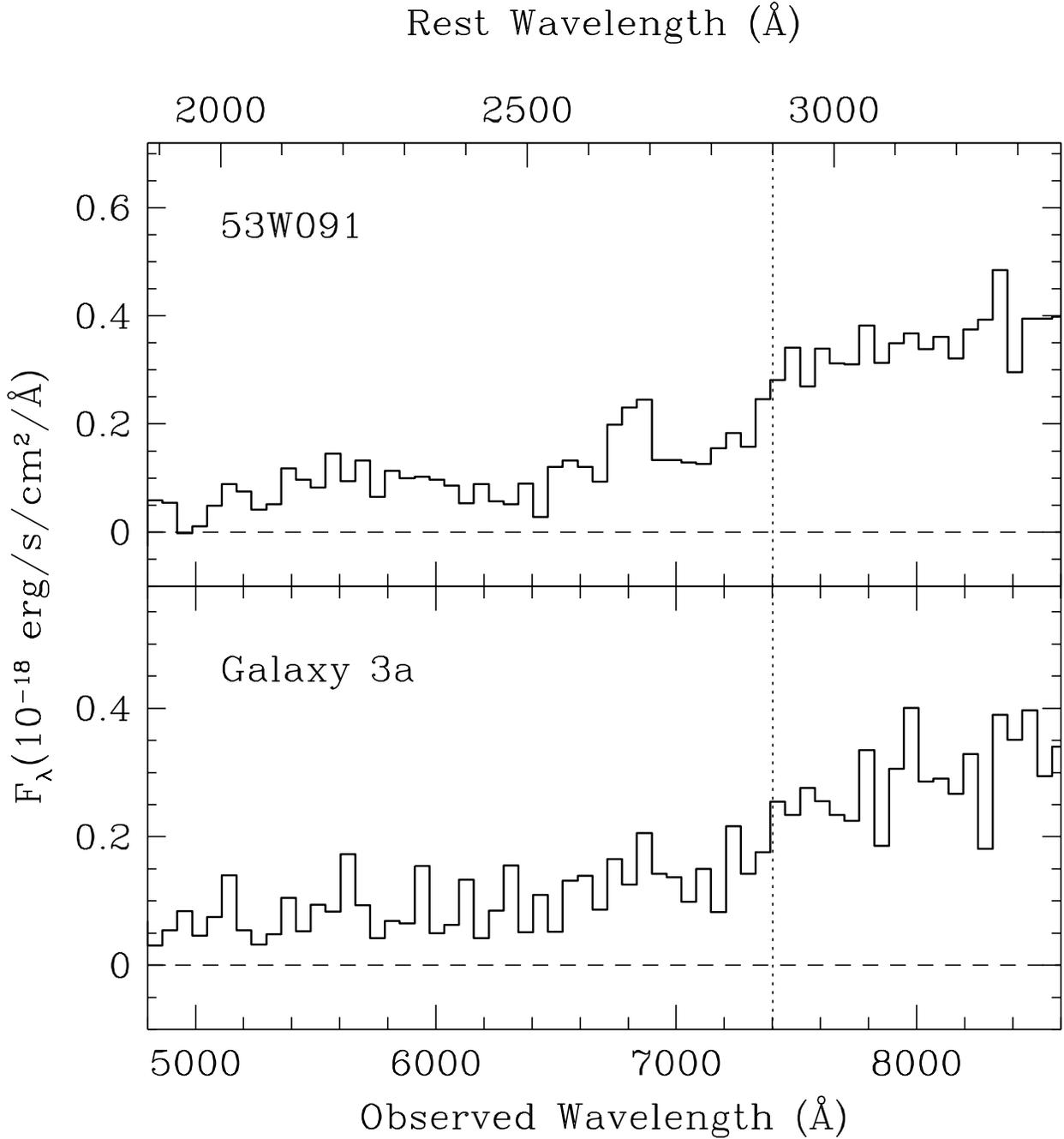
\figcaption{Spectra of LBDS~53W091 (shifted) and galaxy 3a plotted in
the observers' frame.  The spectra have been averaged in 25-pixel
bins.  The rest wavelength is indicated along the upper abscissa for a
redshift of $z=1.552.$ The 2900\AA\ discontinuity apparent in both
objects.  We interpret galaxy 3a to be a faint companion to LBDS~53W091
with both similar age and redshift.  \label{galaxy3a}}


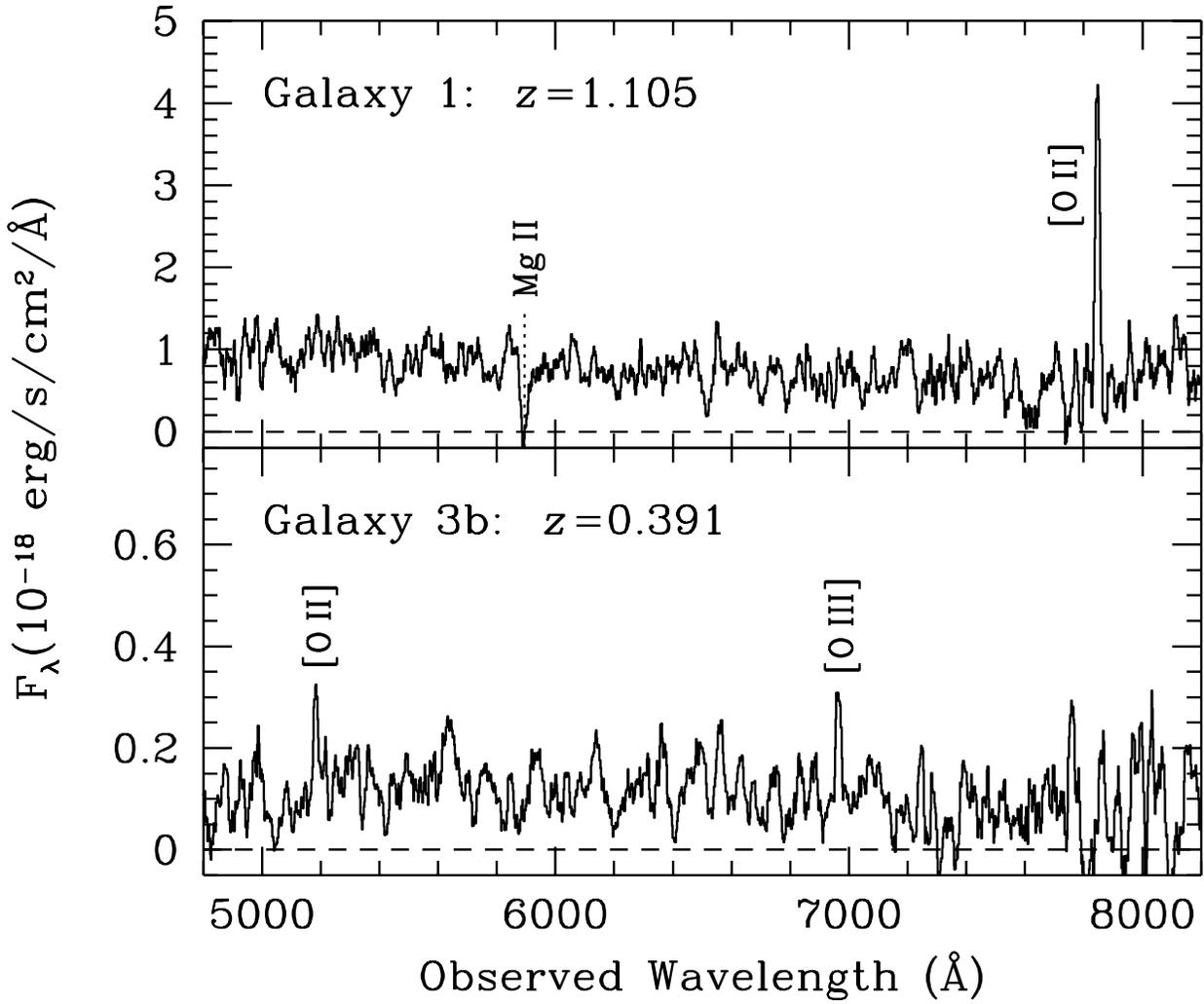
\figcaption{Spectra of the blue emission line galaxies labelled 
``1'' (upper panel) and ``3b'' (lower panel) in Figure 3. The spectra are
plotted in the observed frame. The parameters
of the emission lines are listed in Table~3. \label{galaxy1}}


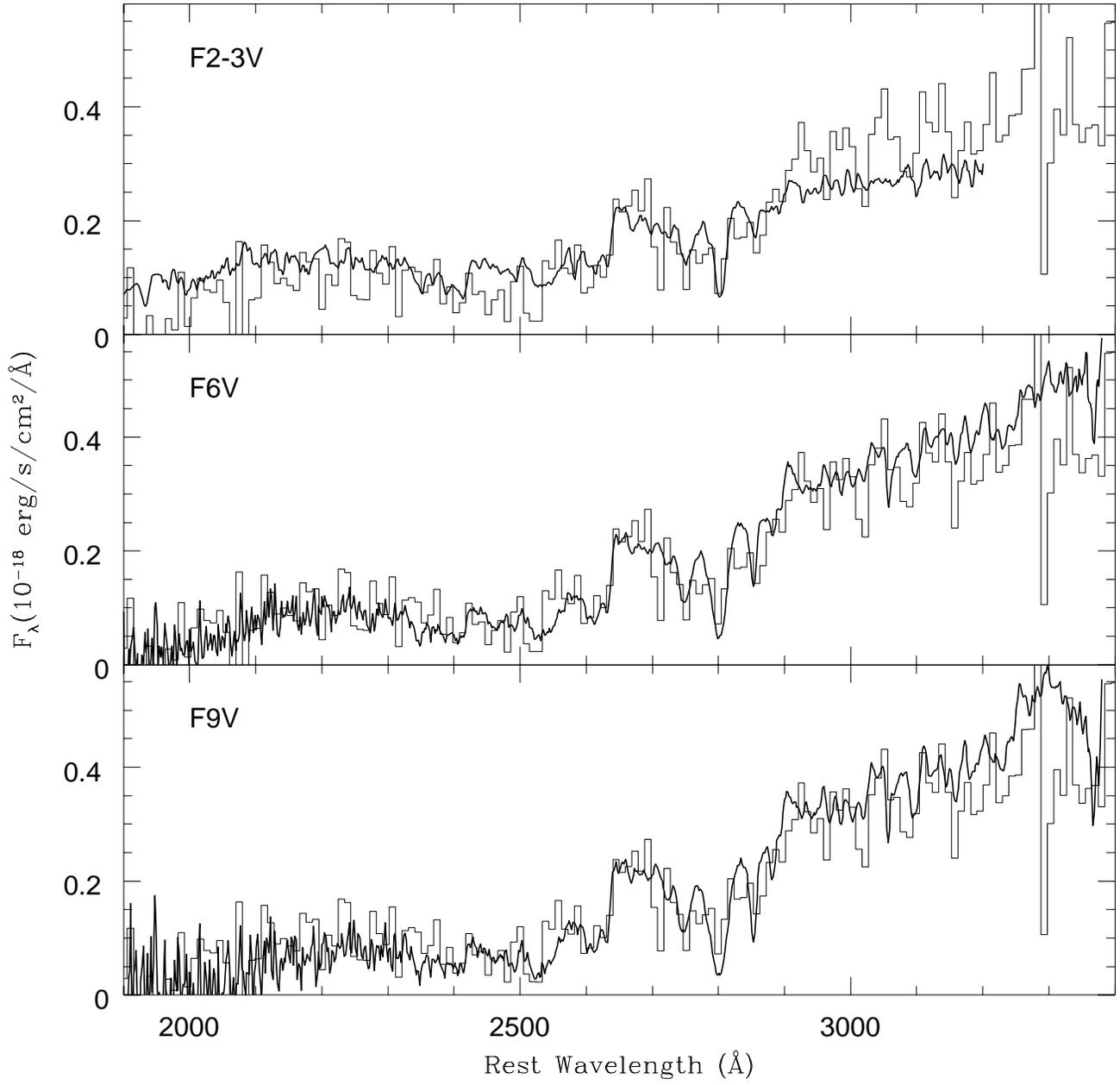
\figcaption{Rest frame spectrum of LBDS~53W091 plotted against scaled
averages of {\it IUE} stars. Note that the spectrum of the average F6V
stellar type the galaxy spectrum almost perfectly. Assuming Solar
metallicity Revised Yale Isochrones, this implies a minimum age just
less than 3 Gyr for the bluest turn--off. \label{iuestar}}


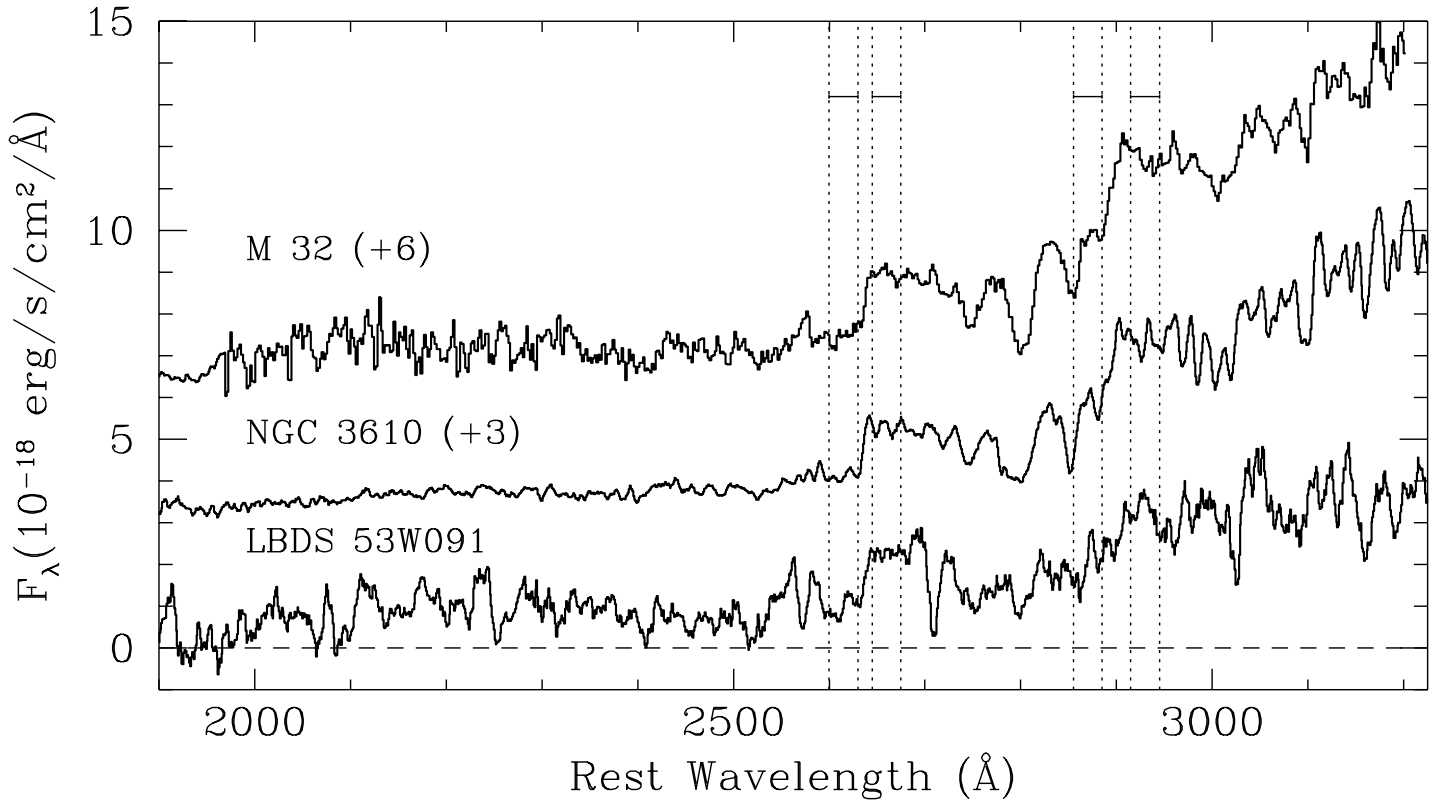
\figcaption{Rest frame spectra of LBDS~53W091 (Keck), M32 ({\it IUE;}
Burstein \etal 1988), and NGC~3610 ({\it HST;} Ferguson, private
communication), where the latter two galaxy spectra have been scaled
and offset.  Note the similarity in the spectral features.  NGC~3610 is
a moderately old nearby elliptical galaxy, with dynamical signs of past
merger activity, and a spectrum slightly stronger--lined than M32.
LBDS~53W091 is slightly bluer, indicating a slightly younger age.  The
horizontal lines indicate the spectral ranges which we use to define
the break amplitudes $B(2640)$ and $B(2900)$, as defined in the text.
\label{m32}}


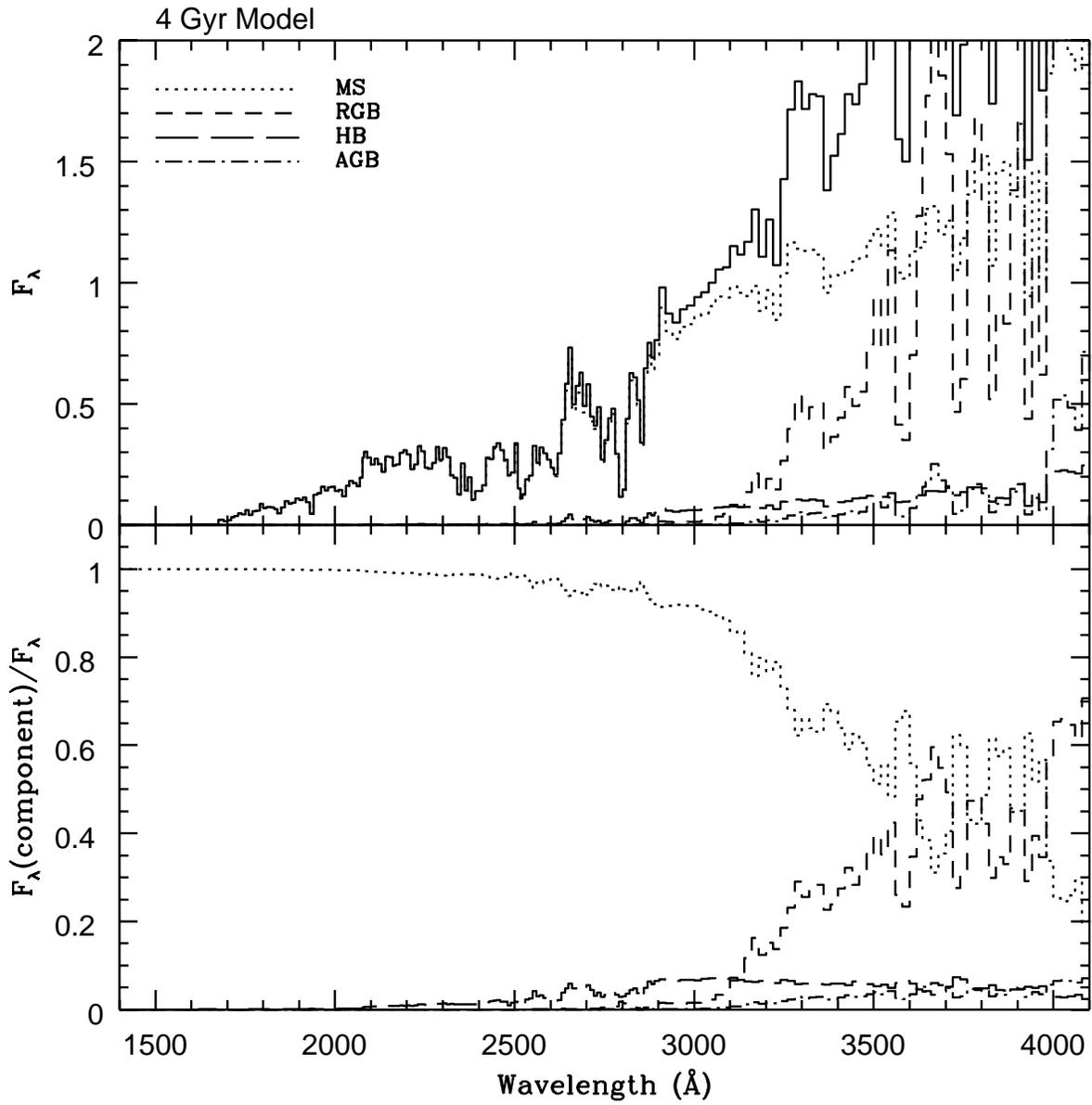
\figcaption{The fractional contribution of different stellar
evolutionary components to the total UV light of an integrated spectrum
at an age of 4~Gyr. The model shown is from the synthesis calculations
of Jimenez \etal (1996).  Note that the main--sequence stars dominate the
flux at $\lambda \simlt 3500$\AA. \label{4gyrstars}}


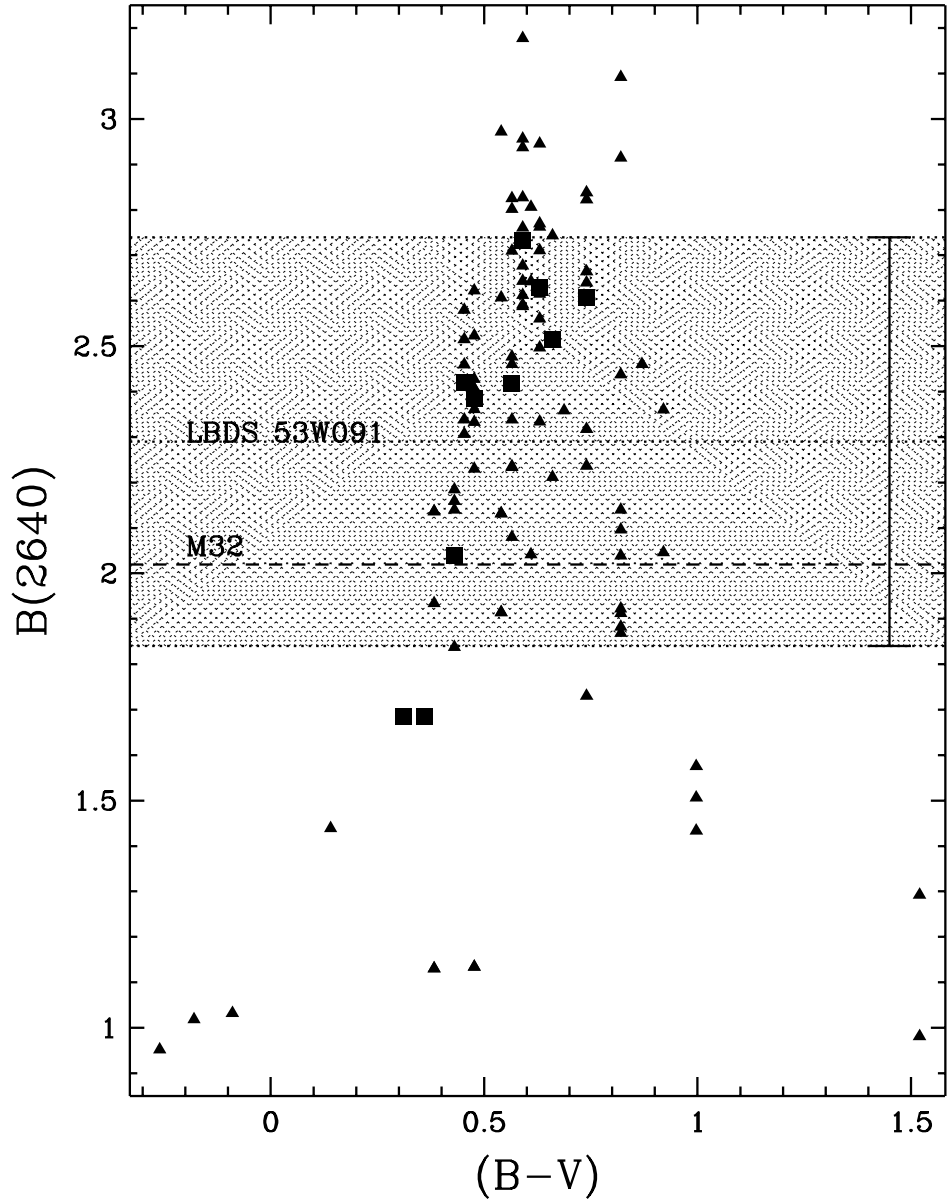
\figcaption{$B(2640)$ break amplitude plotted against $(B - V)$ for
main--sequence stars observed by {\it IUE}.  The solid triangles
represent individual stars, and the solid squares are measured from
average spectra of stars with similar spectral types.  Horizontal lines
indicate the value of this break amplitude measured for the galaxies
M32 and LBDS~53W091.  The large scatter in the strength of this break
with spectral type only provides a lower limit to the color of the UV
bright population of LBDS~53W091, and implies a main sequence turn--off
color of $(B - V) > 0.4$.  \label{b2640star}}


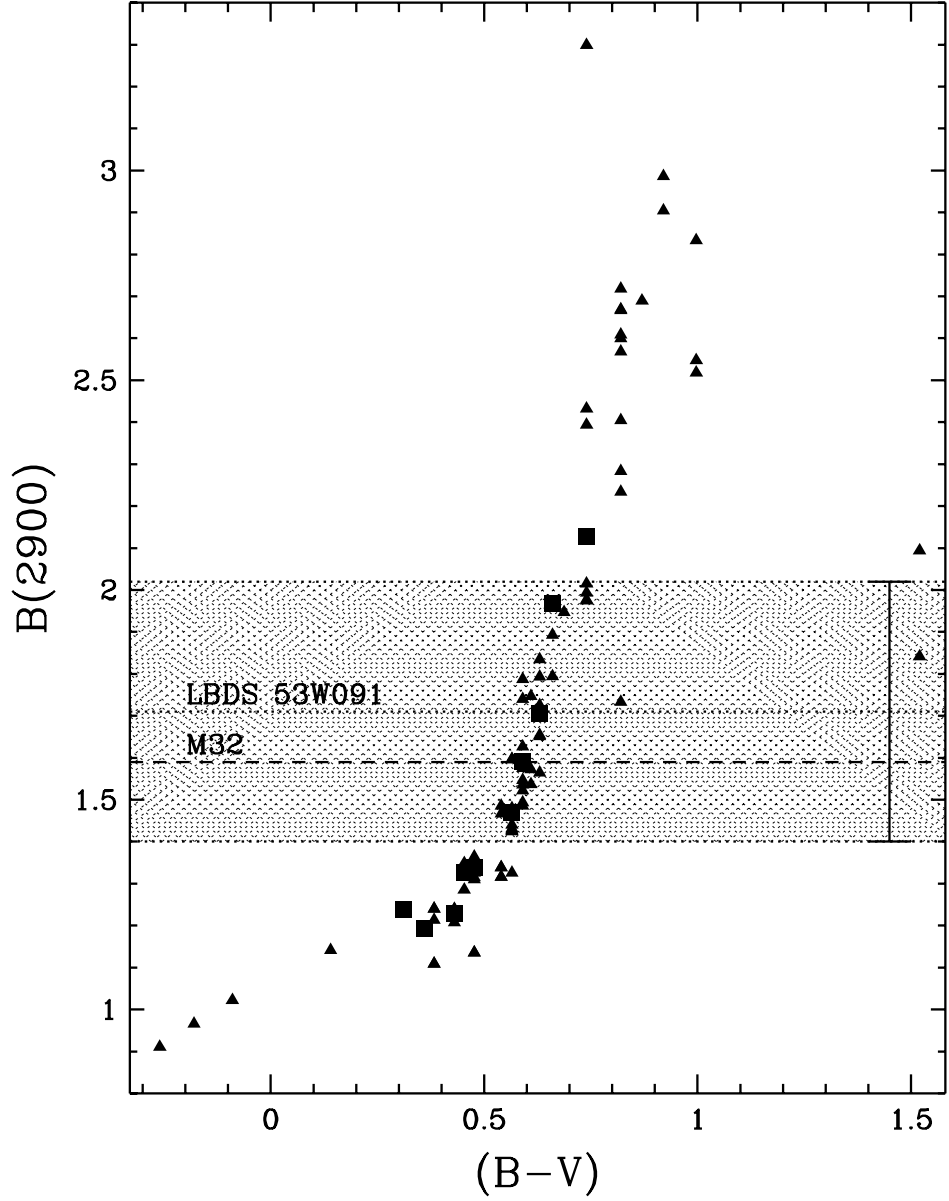
\figcaption{$B(2900)$ break amplitude plotted against $(B - V)$ for
{\it IUE} main--sequence stars.  The symbols are the same as in Figure
11.  Horizontal lines indicate the value of this break amplitude
measured for the galaxies M32 and LBDS~53W091.  This comparison
provides a tighter constraint than the $B(2640)$ break in the previous
figure, and implies that the dominant UV population in LBDS~53W091 has
a main sequence turn--off color of $0.55 < (B - V) < 0.75.$
\label{b2900star}}


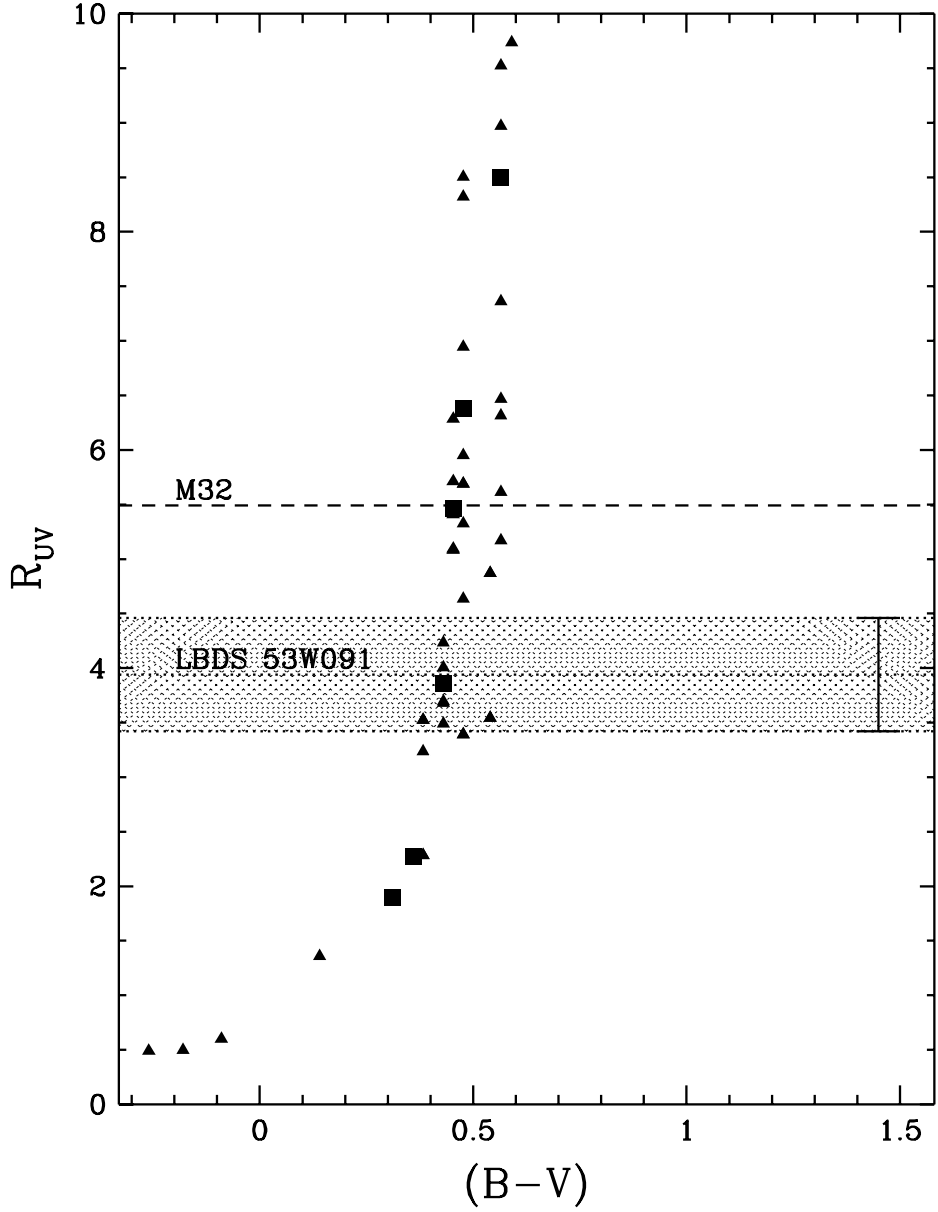
\figcaption{UV color index $R_{UV}$ plotted against $(B - V)$ for {\it
IUE} stars.  The symbols are the same as in Figure 11.  Horizontal
lines indicate the value of this break amplitude measured for the
galaxies M32 and LBDS~53W091.  The spectrum of LBDS~53W091 is
consistent with a main sequence turn--off color of $0.45 < (B - V) <
0.55$, and is therefore consistent with the age estimates determined
from the $B(2640)$ and $B(2900)$ spectral breaks.  \label{iueRUV}}


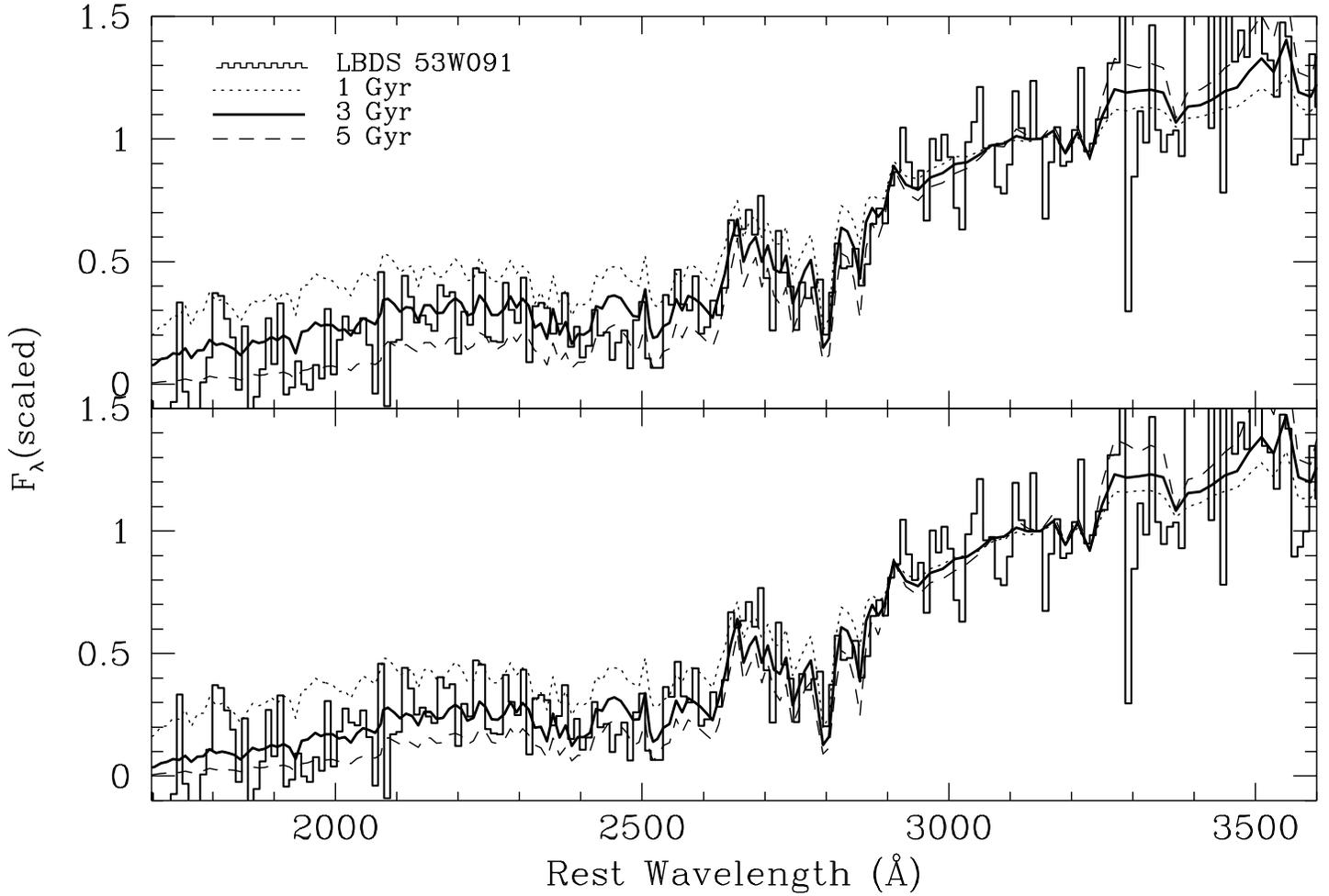
\figcaption{Synthetic spectra at ages of 1, 3 and 5~Gyr from the Solar
metallicity evolutionary models of Jimenez \etal (1996) compared with
the observed spectrum of LBDS~53W091. The upper panel shows the 
main--sequence models, and the lower panel shows the ``full'' models of
Jimenez (1996) (see text).  The flux (in units of $F_\lambda$) is
arbitrarily scaled to unity at 3150\AA\ for all spectra.  Models with
ages less than 3 Gyr are inconsistent with LBDS~53W091.  
\label{raulmodels}}


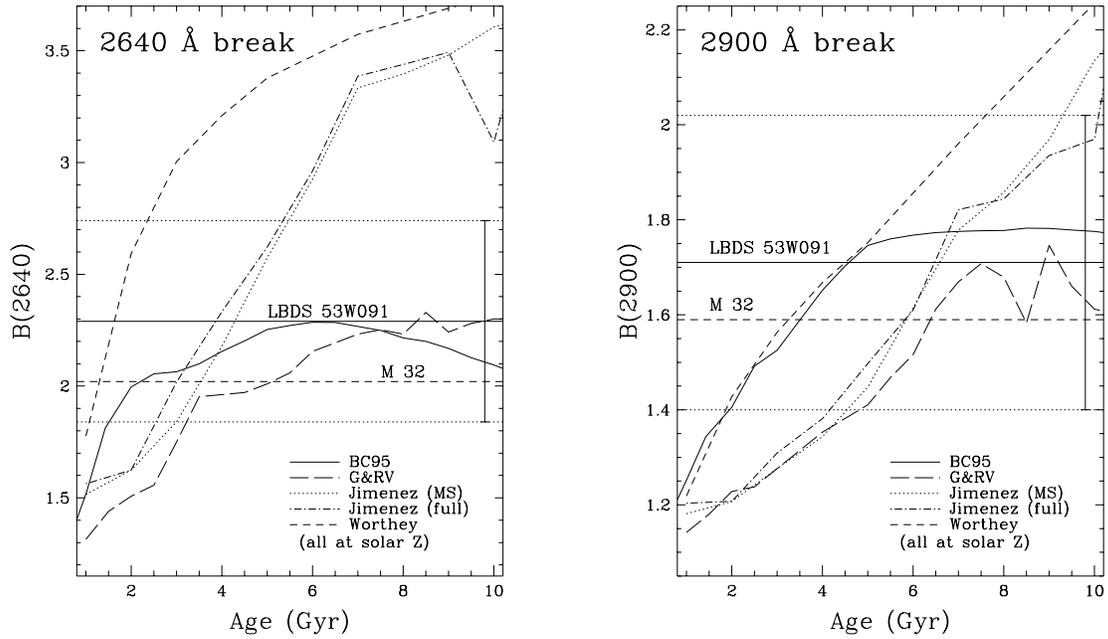
\figcaption{$B(2640)$ (a) and $B(2900)$ (b) spectral discontinuities for
several models, as indicated in the figure.  Horizontal lines are the
measured break amplitudes for LBDS~53W091 and M32, as labelled, where
the formal 1$\sigma$ error  on the value for LBDS~53W091 is also
indicated.  Note the bimodal distribution of model predictions of the
break amplitudes:  models which use Kurucz theoretical stellar spectra
in the UV (Jimenez and Worthey) have break amplitudes which continually
rise, while models which use observed {\it IUE} stars to form the
spectral library (BC95 and G\&RV) asymptote at a break amplitudes of
$B(2640) \approx 2.2$ and $B(2900) \approx 1.7.$ \label{breaksall}}


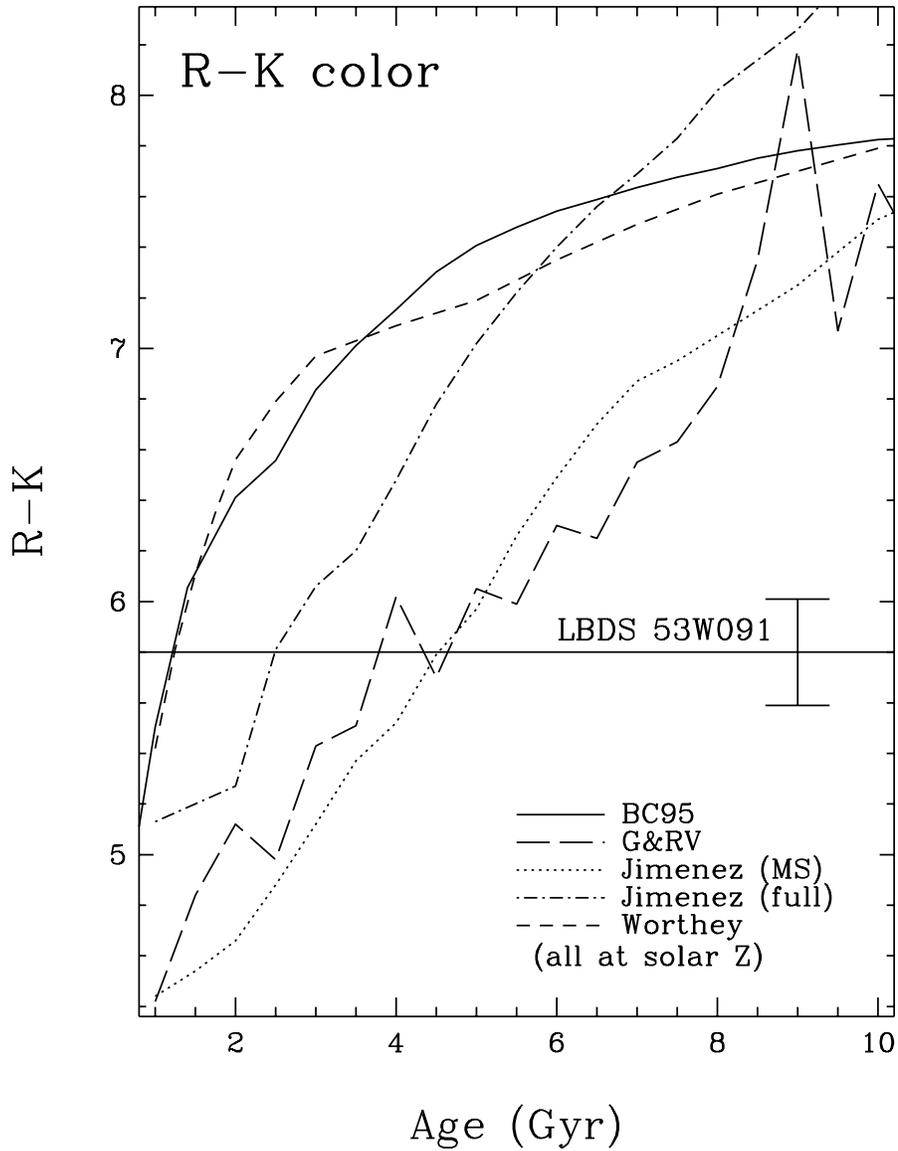
\figcaption{{\it R--K} color for several models, as indicated in the
figure. The models of
Worthey and BC95 imply a very young age for LBDS~53W091, ages which are
inconsistent with the UV spectrum of the galaxy. The models of G\&RV
and the simple main--sequence model, both of which omit AGB stars from
the spectral library (though G\&RV have red subgiants and giants) imply
an age around 4 Gyr for the galaxy.  \label{rk}} 


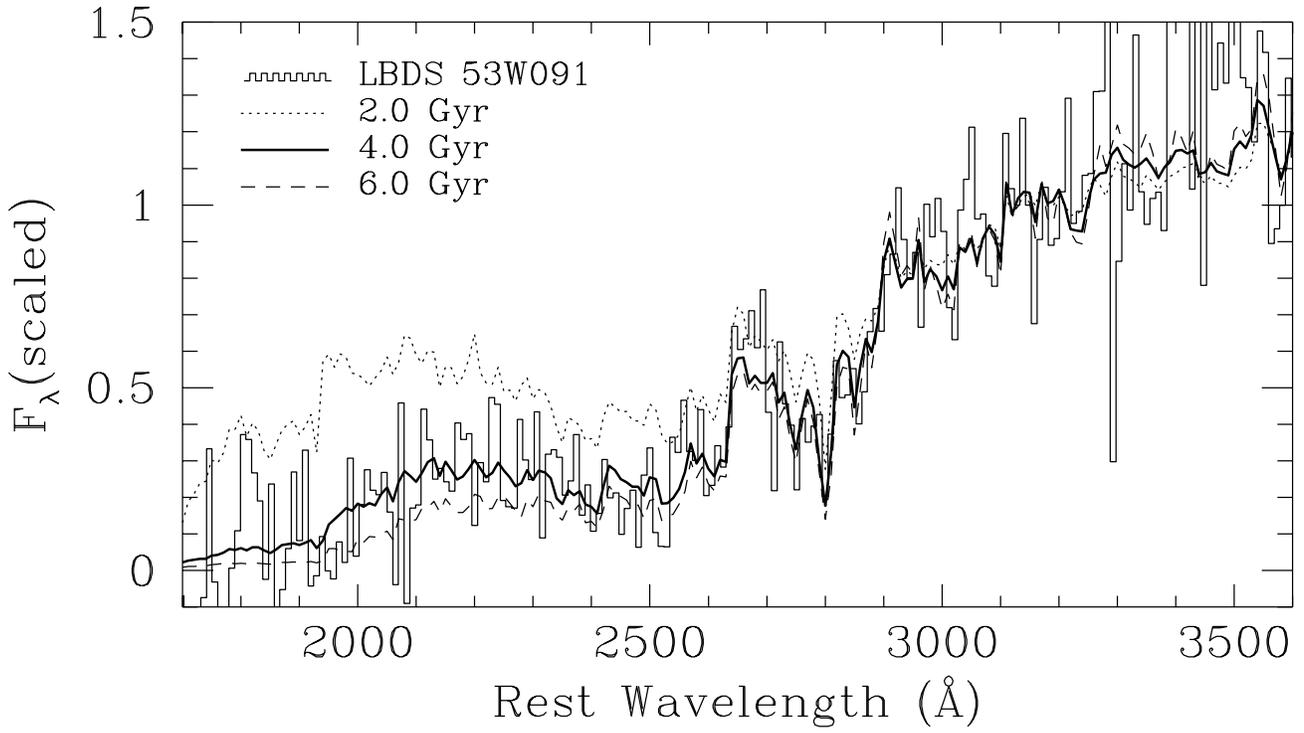
\figcaption{The models of Guideroni and Rocca--Volmerange compared with
the observed spectrum of LBDS~53W091.  The flux is arbitrarily scaled
to unity at 3150\AA\ for all spectra. Models with ages $\simlt 3$~Gyr are
inconsistent with LBDS~53W091. \label{paris}}


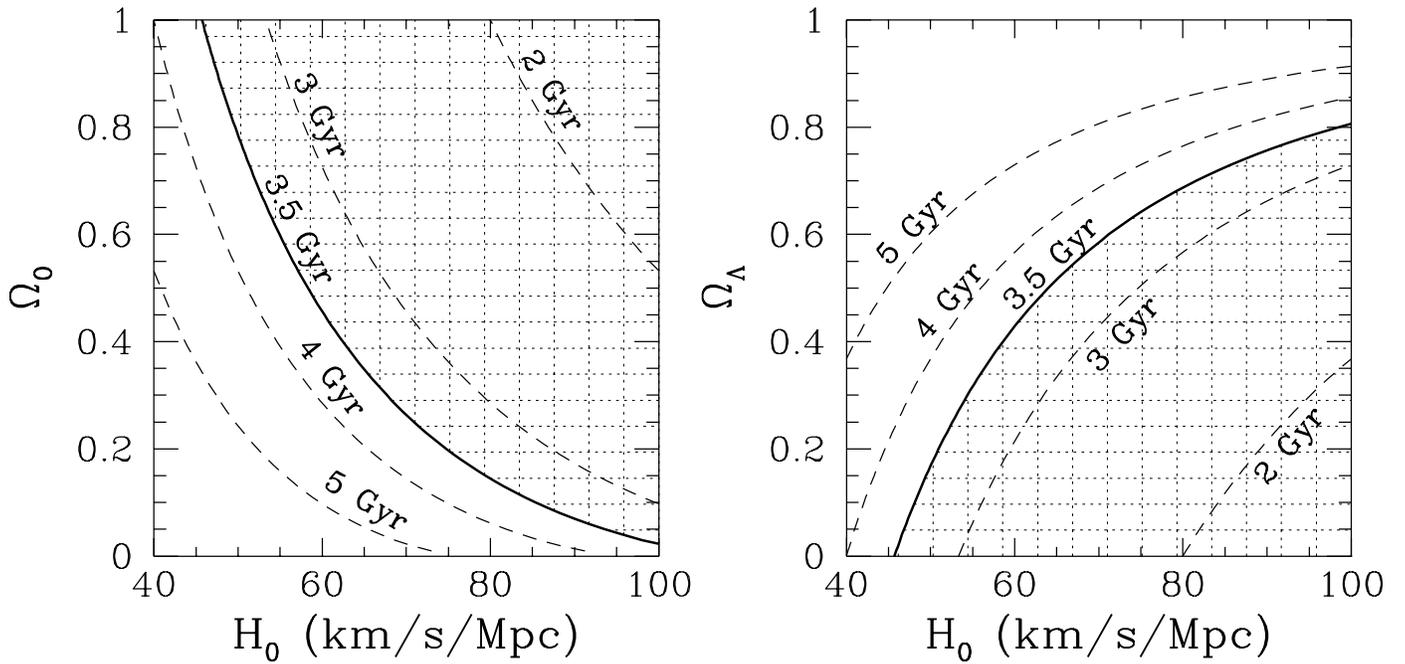
\figcaption{Constraints on the cosmological parameters \hnot,
$\Omega_0,$ and $\Omega_\Lambda$ derived from the age of LBDS~53W091.
We plot the age of the Universe at a redshift of $z=1.552$ for a range
of cosmological parameters.  Models in the left panel assume $\Lambda =
0.$ Models in the right panel assume a flat universe with a
cosmological constant, \ie\ $\Omega_0 + \Omega_\Lambda = 1.$ By
virtue of LBDS~53W091 being older than 3.5 Gyr at this redshift, the
hatched regions of parameter space are forbidden. \label{cosmoplot}}

\end{document}